\documentclass[aps,pre,twocolumn,amsmath,showpacs,superscriptaddress]{revtex4}

\usepackage{amsmath}
\usepackage{amsbsy}
\usepackage{amscd}
\usepackage{amsopn}
\usepackage{amstext}
\usepackage{amsxtra}
\usepackage{epsfig}

\def\kp{{k^\prime}}
\def\wP{{\widetilde{P}}}
\def\wz{{\widetilde{z}}}
\def\lesssim{\mathrel{\hbox{\rlap{\hbox{\lower4pt\hbox{$\sim$}}}\hbox{$<$}}}}

\def\lb{\left}  
\def\rb{\right}

\def\nn{\nonumber}

\begin{document}

\title{Analytical results for bond percolation and k-core sizes on clustered networks}
\author{James P. Gleeson}
\author{Sergey Melnik}
\affiliation{Department of Mathematics \& Statistics, University of Limerick, Ireland.}


\begin{abstract}
An analytical approach to calculating bond percolation thresholds,
sizes of $k$-cores, and sizes of giant connected components on
structured random networks with non-zero clustering is presented.
The networks are generated using a generalization of Trapman's [P.
Trapman, Theor. Pop. Biol. {\bf 71}, 160 (2007)] model of cliques
embedded in tree-like random graphs. The resulting networks have
arbitrary degree distributions and tunable degree-dependent
clustering. The effect of clustering on the bond percolation
thresholds for networks of this type is examined and contrasted
with some recent results in the literature. For very high levels
of clustering the percolation threshold in these generalized
Trapman networks is increased above the value it takes in a
randomly-wired (unclustered) network of the same degree
distribution. In assortative scale-free networks, where the
variance of the degree distribution is infinite, this clustering
effect can lead to a non-zero percolation (epidemic) threshold.
\end{abstract}

\maketitle

\section{Introduction}
There has been considerable recent interest in the study of random
network models, with a view  to understanding the structure and
dynamics of the Internet, citation networks, and other social,
biological and technological networks; see the
reviews~\cite{Newman03a, Dorogovtsev08, Boccaletti06,
Dorogovtsev03} and references therein. The degree distribution
$P_k$ is a fundamental quantity of interest in these studies; here
$P_k$ is defined as the probability that a randomly chosen node
(vertex) in the network has $k$ neighbors. Random networks with a
specified $P_k$ may be generated using the so-called
\emph{configuration model}~\cite{Newman01a}, which randomly links
pairs of nodes to give the correct degree distribution. The
properties of networks generated in this manner are now well
understood, with analytical results relying on the fact that such
networks can be approximated very accurately by tree-like graphs
(provided that $P_k$ decays sufficiently rapidly for large
$k$~\cite{Newman01a, Burda04, Bianconi08}).

However, most real-world networks are not tree-like, since the density of cycles (loops) of length three in such networks is non-zero, whereas this quantity vanishes (in the limit of infinite network size) for the configuration model. The \emph{local clustering coefficient} for a node $A$ is defined as the fraction of pairs of neighbors of node $A$ which are also neighbors of each other~\cite{Watts98}. The \emph{degree-dependent clustering} $c_k$ is the average of the local clustering coefficient over the class of all nodes of degree $k$~\cite{Serrano06a,Vazquez02}. Because analytical results are difficult to obtain for networks containing loops, the question of how models incorporating both $P_k$ and non-zero $c_k$ (taken, for example, from real-world network data) differ in structure and dynamics from corresponding randomly-wired networks (where $c_k\to 0$) remains of considerable interest.

The bond percolation problem on networks depends strongly on the
structure of the underlying graph, and also has several important
applications. The problem  may be stated as follows: each edge of
the network graph is visited once, and \emph{damaged} (deleted)
with probability $1-p$ (the quantity $p$ is the \emph{bond
occupation probability}). The size of the \emph{giant connected
component} (GCC) of the graph is clearly zero for $p=0$ but
becomes nonzero at some critical value of $p>0$: this critical
value of $p$ is termed the \emph{bond percolation threshold}
$p_c$. The bond percolation problem has applications in
epidemiology, where $p$ is related to the average transmissibility
of a disease and the GCC represents the size of an epidemic
outbreak, and in the analysis of technological networks, where the
resilience of a network to the random failure of links is
quantified by the size of the GCC~\cite{Serrano06b}. The
percolation threshold and the GCC size may be determined
analytically for configuration model networks~\cite{Callaway00}.

A number of investigations into the effects of clustering on bond
percolation have also been undertaken. Newman~\cite{Newman03b}
introduced a bipartite graph model of highly clustered networks,
and examined an example of a network in which the existence of
clustering decreases the percolation threshold from its value in
an unclustered network, see also~\cite{Britton07}. Serrano and
Bogu\~{n}\'{a}~\cite{Serrano06c, Serrano06a, Serrano06b} make a
detailed analysis of the interdependence of clustering and
correlations. They distinguish between two types of clustered
networks: those with average clustering $c_k$ of $k$-degree nodes
less than $1/(k-1)$, termed \emph{weakly clustered}, and those
with $c_k>1/(k-1)$, termed \emph{strongly clustered}. The boundary
$c_k=1/(k-1)$ represents the largest value of clustering
achievable without inducing degree-degree correlations in the
network. Using approximate analytical methods for the weak
clustering cases and numerical simulations~\cite{Serrano06b} for
some strongly clustered networks, they compare the bond
percolation threshold to the value it would have for an
unclustered network with the same degree distribution. Their
general conclusion is that weak clustering increases the
percolation threshold above its unclustered value, while strong
clustering decreases the threshold. The latter conclusion is
consistent with the example examined by Newman~\cite{Newman03b}.
On the other hand, it has been pointed out in the epidemiological
literature~\cite{Eames08, Miller08} that in clustered networks
infection tends to be confined within highly connected groups, and
so sufficient clustering should increase the epidemic
(percolation) threshold.

Trapman~\cite{Trapman06, Trapman07} recently introduced a model of
clustering in  structured graphs based on embedding cliques
(complete subgraphs) within a random tree structure. He uses this
model to analytically determine epidemic thresholds on networks
with non-zero clustering. In Trapman's model networks, the
degree-dependent clustering $c_k$ is of the form $c_k \propto
(k-2)/k$ for all $k\ge 3$. In particular, $c_k$ increases with
increasing degree $k$, which is contrary to the typically
decreasing behavior $c_k \sim k^{-1}$ for large $k$ seen in
real-world networks~\cite{Dorogovtsev02,Ravasz03}. In this paper
we generalize the Trapman construction to allow for more general
$c_k$ dependence on $k$ (see equation~(\ref{cf}) below), with a
view to matching to the degree-dependent clustering of real-world
networks. As shown in section~\ref{s:pc}, this generalization
leads to clustered networks in which the bond  percolation
threshold may be either larger or smaller than the threshold in a
randomly-wired (configuration model) network with the same degree
distribution $P_k$. Furthermore, we develop methods
from~\cite{Gleeson08a} to give analytical results for the GCC
(epidemic) size on clustered networks. We also demonstrate the
adaptability of these methods by calculating the sizes of
$k$-cores on clustered networks. The $k$-core of a network is the
largest subgraph whose nodes have degree at least
$k$~\cite{Bollobas84,Goltsev06}; study of $k$-core decompositions
gives insights into the topology of interconnected parts of
real-world networks such as the Internet~\cite{Carmi07}.
Analytical results for $k$-core sizes have been found for
configuration model networks~\cite{Dorogovtsev06} and on tree-like
random graphs with degree-degree correlations~\cite{Gleeson08a},
but both these cases assume zero clustering in the network. Very
recently, alternative models for random graphs with clustering
have been published \cite{Newman09,Gleeson09a}, but these examine
only the bond percolation problem.

The layout of the paper is as follows. The generalization of
Trapman's algorithm for generating clustered networks is described
in section~\ref{s:GenClustNet}. In section~\ref{s:pc} we  examine
the transition point for bond percolation on such clustered
networks, and show that clustering may either increase or decrease
the epidemic threshold. Comparisons are drawn with results using
data for some real-world networks. Section~\ref{s:GCC} describes
an analytical approach to calculating the size of the giant
connected component (the epidemic size), and the method is
extended in section~\ref{s:kcore} to yield $k$-core sizes.
Finally, conclusions are drawn in section~\ref{s:Concl}.

\section{Generating the clustered network} \label{s:GenClustNet}
\begin{figure}
\centering
\includegraphics[width=0.47\columnwidth]{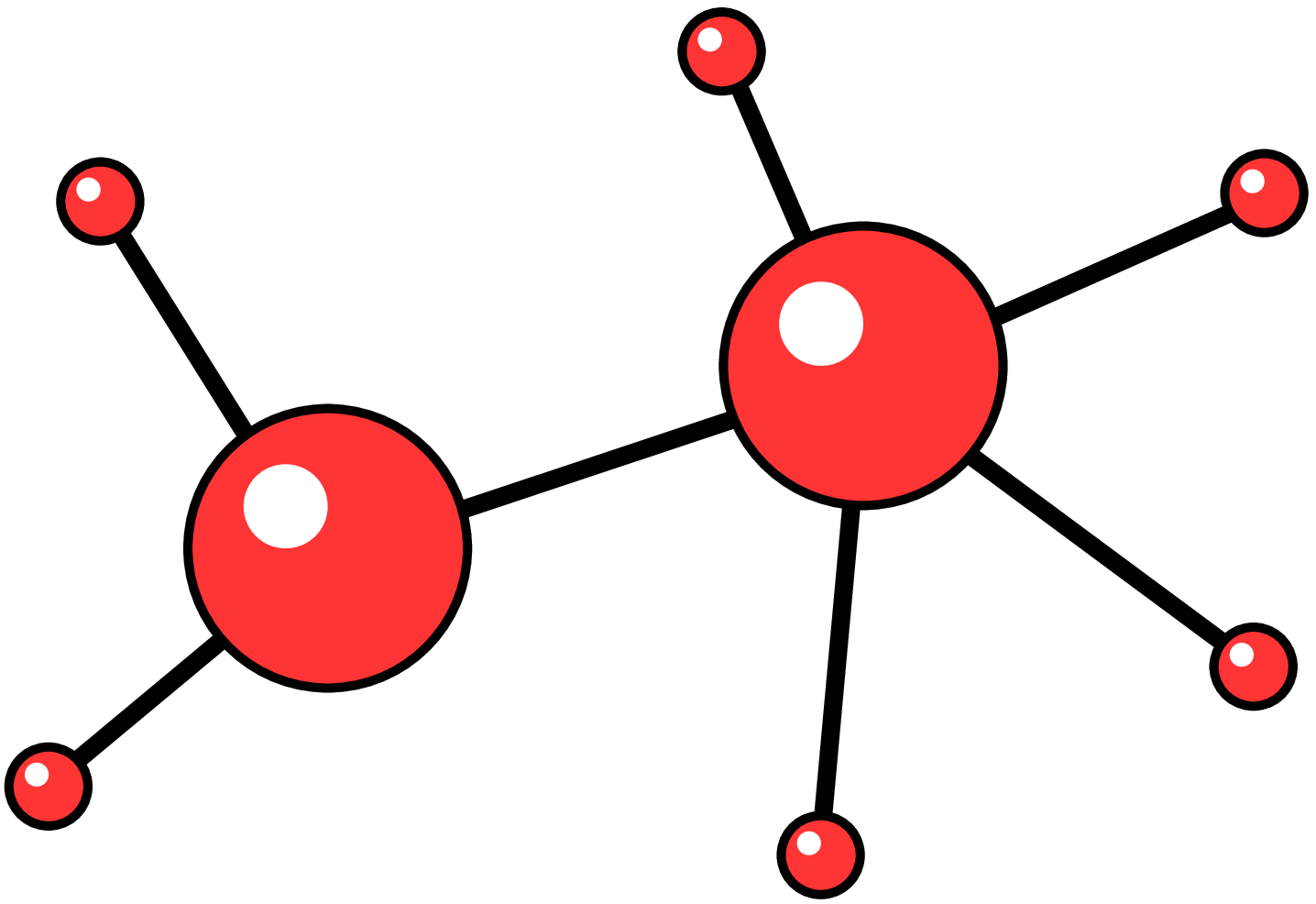}
\hspace{0.3cm}
\includegraphics[width=0.47\columnwidth]{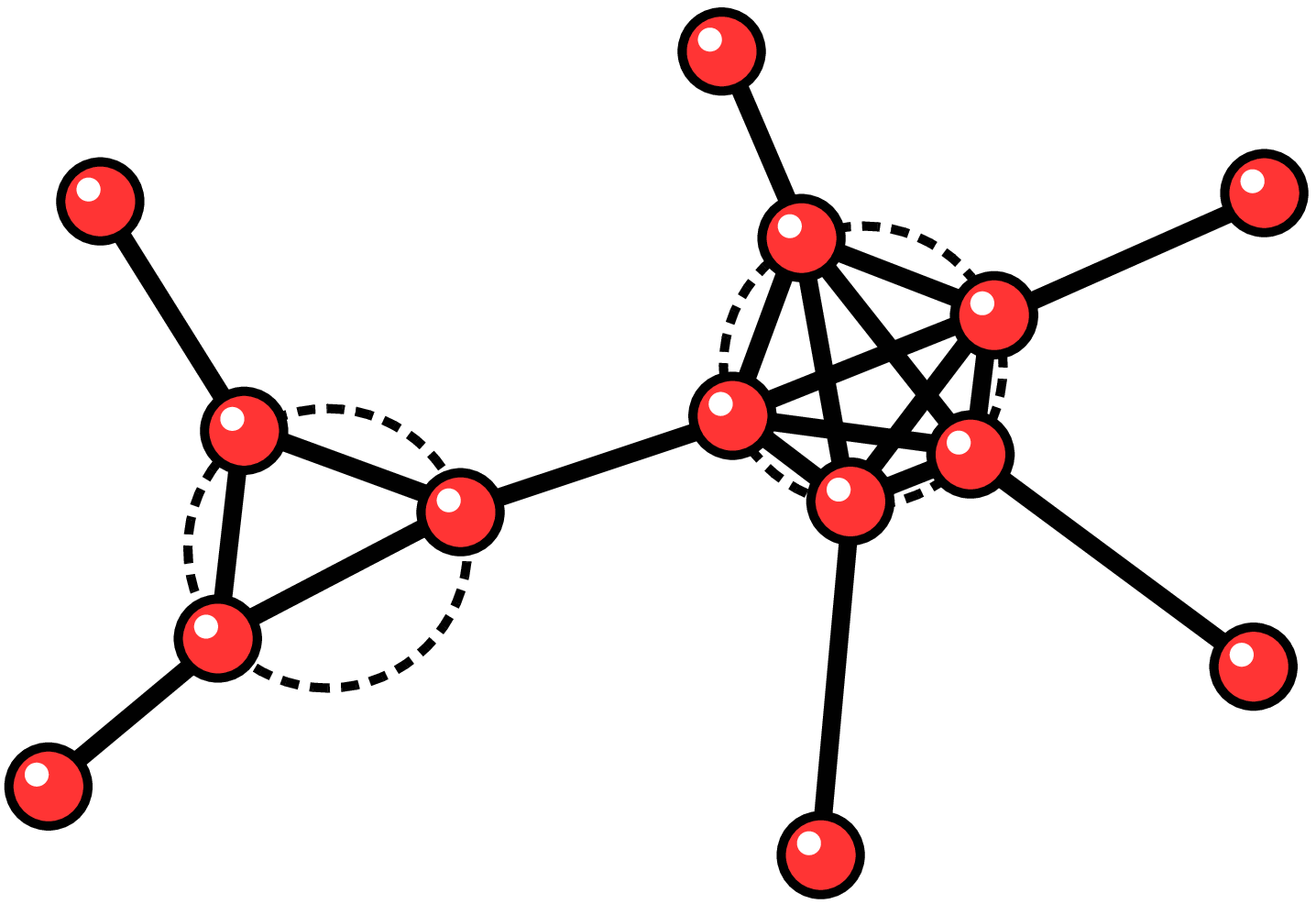}
\put(-240,70){\bf(a)} \put(-110,70){\bf(b)} \caption{(Color
online) (a) Graph of super-individuals which consists of two
household nodes and six bachelor nodes. (b) Graph of individuals
which is generated from (a) by expanding households into
$k$-cliques of individual nodes.} \label{f:SuperIndiv}
\end{figure}

Here we describe an algorithm based on that
of~\cite{Trapman06,Trapman07} which generates structured random
networks with arbitrary degree distributions $P_k$ and with high
clustering. The algorithm can be written in three steps, as
follows:
 \begin{itemize}
 \item[(i)] An uncorrelated random network is created using the configuration model in the standard way (connecting stubs at random). This network,
 which we call the \emph{super-graph},
 has a finite-variance degree distribution $\widetilde{P}_k$, related to the desired distribution  $P_k$ of the final network by equation (\ref{pp}) below. The nodes of this super-graph are called
 \emph{super-individuals}.
 \item[(ii)]
A fraction $g_k$ of all $k$-degree super-individuals (for $k\ge 3$) are tagged as \emph{households}. This tagging does not affect the random linking of the configuration model in any way, but is used in the next step of the algorithm. The untagged super-individuals will be referred to as \emph{bachelors}. Figure~\ref{f:SuperIndiv}(a) shows an example of a super-graph, with two households (drawn as larger nodes) and six bachelors.
 \item[(iii)]
Taking the tagged super-graph of step (ii) as input, we generate the \emph{individuals graph}, in which each node  represents a single individual. Each super-individual (of degree $k$ say) which is tagged as a household is expanded into a $k$-clique of individual nodes. Thus each household in the super-graph is replaced in the individuals graph by $k$ individuals of degree $k$, all of whom are linked to each other, and each of which has one neighbor outside his own household (see Figure~\ref{f:SuperIndiv}(b)). Each bachelor in the super-graph becomes an individual in the individuals graph. When all super-individuals have been replaced in this way we have generated the \emph{individuals graph} with degree distribution $P_k$ and the algorithm concludes.
\end{itemize}
Let $\widetilde N$ be the total number of super-individuals in the super-graph of step (i). When $\widetilde N$ is sufficiently large,  there are approximately $\widetilde N \wP_k$ super-individuals of degree $k$ in the network. The bachelors among these become $\widetilde N\wP_k(1-g_k)$ individual nodes of degree $k$, while the households of degree $k$ are expanded to $\widetilde N \wP_k g_k k$ individuals grouped into $k$-cliques. Letting $N$ denote the total number of individuals, we sum over all degree classes to obtain the relation
\begin{equation}
N = \widetilde N \sum_k \wP_k\lb(1-g_k + k \,g_k\rb).
\label{NN}
\end{equation}
Note that taking the limit $\widetilde N \to \infty$ therefore implies $N \to \infty$, and vice versa.

It is convenient to introduce the fraction $f_k$ of $k$-degree nodes in the individuals graph which are members of a $k$-clique. This fraction is related to the fraction $g_k$ of $k$-degree super-individuals who were tagged as households in step (ii) of the algorithm:
\begin{equation}
g_k = \frac{f_k}{f_k+k-k f_k} \quad \iff \quad
f_k = \frac{k g_k}{1-g_k+k g_k}. \label{fg}
\end{equation}
In terms of $f_k$ we have the following relation between the degree distributions $\widetilde{P}_k$ and $P_k$ of the super- and individuals graphs respectively:
\begin{equation}
 \widetilde{P}_k = \frac{P_k\lb(1-f_k+f_k/k\rb)}{\sum_{\kp=0}^\infty
 P_{\kp}\lb(1-f_\kp+f_\kp/\kp\rb)}.
 \label{pp}
\end{equation}
Trapman's original model~\cite{Trapman06,Trapman07} constrains the degree distribution of bachelors within the super-graph to match the distribution $P_k$ of the individuals graph. This case corresponds to choosing $f_k$ to be independent of $k$, i.e., $f_k=F$ for constant $F$. As we show in subsequent sections, many new phenomena arise when $f_k$ depends on $k$; we will refer to this case as the \emph{generalized Trapman model}.

The degree-dependent clustering  coefficients $c_k$ in the final,
individuals graph may be calculated by noting that each $k$-degree
individual is either a member of a single $k$-clique (with
probability $f_k$) or is a member of no clique (with probability
$1-f_k$). Since each node in a $k$-clique has clustering level
$(k-2)/k$ and nodes connected using the configuration model have
effectively zero clustering level in the $\widetilde N\to\infty$
limit (and assuming $\widetilde{P}_k$ has finite variance), the
final average clustering for the $k$-degree nodes in the
individuals graph may be written as
\begin{equation}
c_k =\frac{ f_k (k-2)}{k}\quad\text{  for }k\ge 3.
\label{cf}
\end{equation}

Thus, given a desired degree distribution $P_k$ and
degree-dependent  clustering coefficients $c_k$ (for $k\ge 3$),
the set of $f_k$ values may be obtained from (\ref{cf}) with the
degree distribution $\widetilde{P}_k$ and fractions $g_k$ for the
super-graph of step (i) of the algorithm following from (\ref{pp})
and (\ref{fg}) respectively. Therefore this algorithm can produce
structured  random graphs with almost any desired level of
clustering (limited only by the constraint from (\ref{cf}) that
$c_k \le (k-2)/k$, to ensure $f_k \le 1$). Moreover, this model
gives analytically tractable results for  a number of dynamical
processes on networks~\cite{Gleeson08a}. Here we shall concentrate
on the bond percolation problem and the calculation of $k$-core
sizes. In this context it is worth noting that our algorithm,
which permits $k$-degree nodes to be members of at most one
$k$-clique, can be viewed as a restricted version of Newman's
bipartite graph model~\cite{Newman03b}. However, unlike Newman's
model, we can specify the degree distribution $P_k$ a priori. As
noted above, our model is also analytically tractable for a
variety of processes beyond percolation. It must be recognized
that the heavily intermittent clustering due to the $k$-cliques
gives a topological structure that may be very different to a
real-world network with the same $P_k$ and $c_k$; nevertheless the
model can give some useful insights into the effect of clustering
on GCC and $k$-core sizes in complex networks.

\section{Bond percolation threshold} \label{s:pc}
\subsection{Calculating $p_c$ in clustered networks}
The giant connected component (GCC) of an infinite graph exists if
$z_2$, the expected number of second neighbors of a random node,
exceeds $z_1$, the expected number of first
neighbors~\cite{Newman01a}. Note both $z_1$ and $z_2$ are
evaluated on the damaged graph, i.e., after a fraction $1-p$ of
the links have been deleted. The lowest value of $p$ for which
$z_2/z_1=1$ therefore defines the bond  percolation threshold
$p_c$. Here we use this criterion to determine the percolation
threshold (epidemic threshold) in the individuals graphs generated
using the algorithm described in section~\ref{s:GenClustNet}.

Note that a giant connected component can exist in the individuals
graph  only if the super-graph also has a GCC. It is therefore
sufficient to determine a condition for the percolation transition
in the super-graph, while correctly taking account of the internal
$k$-clique structure of the super-individuals which are tagged as
households.

The expected number of first neighbors in the damaged
 super-graph is $\wz_1 = p \wz$, where $\wz=\sum k\wP_k$ is the  mean degree of the undamaged super-graph. To determine the expected number of second neighbors $\wz_2$ in the damaged super-graph, we first choose a super-individual at random. On average, this super-individual has $\wz_1$ first neighbors, with a  given first neighbor being of degree $k$ with probability $k \wP_k/\wz $ \cite{Dorogovtsev03}. If this first neighbor is a bachelor (which occurs with probability $1-g_k$) then it connects on average to $(k-1)p$ super-individuals other than the original. If it is a household (with probability $g_k$) then the connections to the $(k-1)p$ further super-individuals may be thwarted by deleted internal links within the $k$-clique of individuals comprising the household.
  Thus household first neighbors connect on average to $D_k(p)$ new neighbors, where $D_k(p)$ is a polynomial in $p$ which may be determined exactly by methods used in~\cite{Newman03b} (see Appendix A), but whose values are bounded by
\begin{equation}
0\le D_k(p) \le (k-1)p.
\label{ineq}
\end{equation}

Combining the cases listed above, we write the expected number of second neighbors in the damaged super-graph as
\begin{equation}
\wz_2 = \wz_1 \sum_{k=1}^\infty \frac{k}{\wz} \wP_k \lb( (1-g_k)(k-1)p + g_k D_k(p)\rb),
\end{equation}
and so the bond percolation threshold $p_c$ is the lowest value of
$p$ for which $\wz_2/\wz_1=1$, i.e. $p_c$ satisfies the polynomial
equation
\begin{equation}
\sum_{k=1}^\infty \frac{k}{\wz} \wP_k \lb( (1-g_k) (k-1)p_c + g_k D_k(p_c)\rb)=1.
\label{orgcond}
\end{equation}
Using equations (\ref{pp}) and (\ref{fg}) this condition may conveniently be expressed in terms of the degree-distribution $P_k$ of the individuals graph, and the fraction $f_k$ of
$k$-degree individuals in cliques:
\begin{align}
\nn \sum_{k=1}^\infty P_k (& k(k-1)p_c-k + \\
&f_k(k-1-k(k-1)p_c + D_k(p_c)))= 0.
\label{GCCcond}
\end{align}
This is a polynomial equation  for the percolation threshold $p_c$, and its solution requires calculation of the $D_k(p)$ functions as specified in  Appendix A. Note that if $f_k=F$, a constant for all $k$, then this reduces to the criterion determined by Trapman's~\cite{Trapman07} equation (14). Of particular interest is the relationship between $p_c$ and the percolation threshold in unclustered (configuration model) random networks with the same degree distribution $P_k$, known to be given explicitly by~\cite{Callaway00}
\begin{equation}
p^\text{rand}_c = \frac{\sum k {P}_k}{\sum k(k-1)
{P}_k}=\frac{\lb<k\rb>}{\lb<k^2\rb>-\lb<k\rb>}.
\label{allord}
\end{equation}
Here we have introduced the angle bracket notation to denote
averaging with respect to the degree distribution $P_k$. In the
remainder of this section we will examine the sign of
$p_c-p_c^\text{rand}$ to determine whether the bond percolation
threshold in the clustered network is greater than, or less than,
the corresponding threshold in an unclustered network with the
same degree distribution.
\begin{figure}
\centering
\includegraphics[width=0.95\columnwidth]{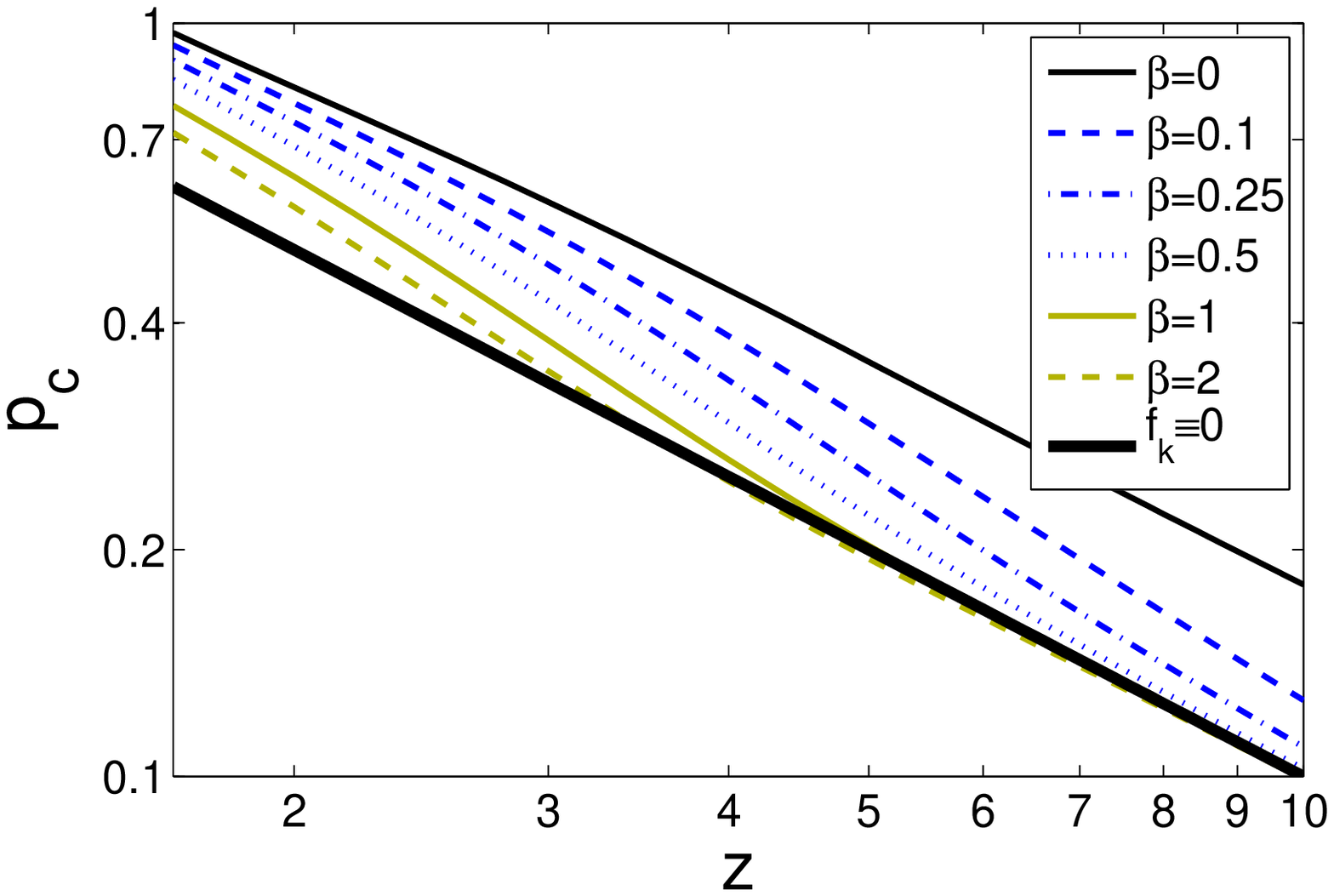}
\put(-240,140){\bf(a)}
\hspace{0.3cm}
\includegraphics[width=0.95\columnwidth]{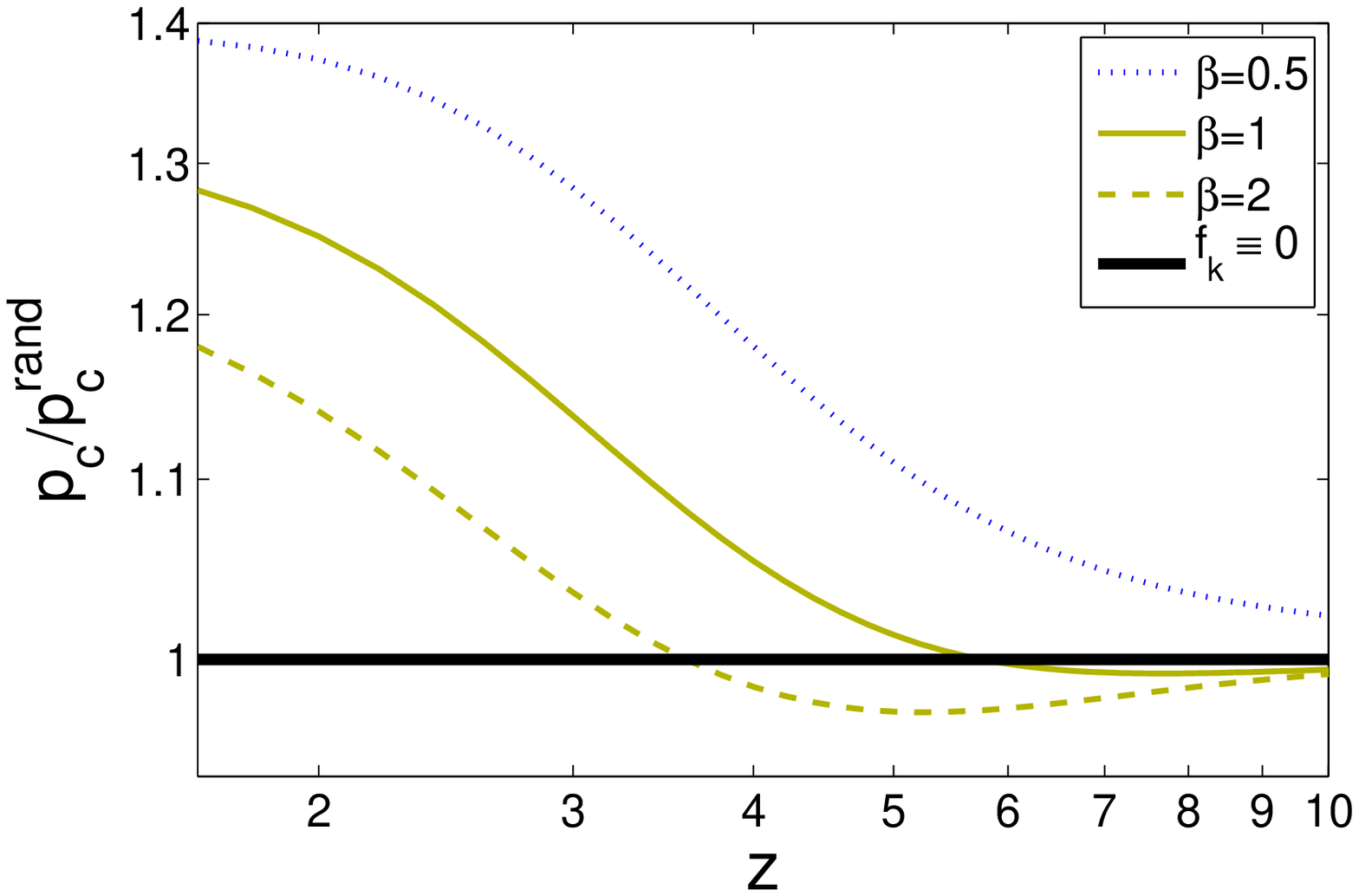}
\put(-240,140){\bf(b)}
 \caption{(Color online)
(a) Bond percolation threshold $p_c$ in clustered Poisson random
graphs with mean degree $z$. The fraction $f_k$ of individuals of
degree $k$ which are members of households ($k$-cliques) is
$f_k=\lb(2/(k-1)\rb)^{\beta}$ with $\beta$ taking values indicated
in the legend. The thick black curve shows the percolation
threshold $p^\text{rand}_c$ in the unclustered ($f_k\equiv 0$)
case. (b) The ratio $p_c/p^\text{rand}_c$ highlights the decrease
in the percolation threshold due to clustering when $\beta$ is 1
or 2.} \label{f:2}
\end{figure}

\begin{figure}
\centering
\includegraphics[width=0.95\columnwidth]{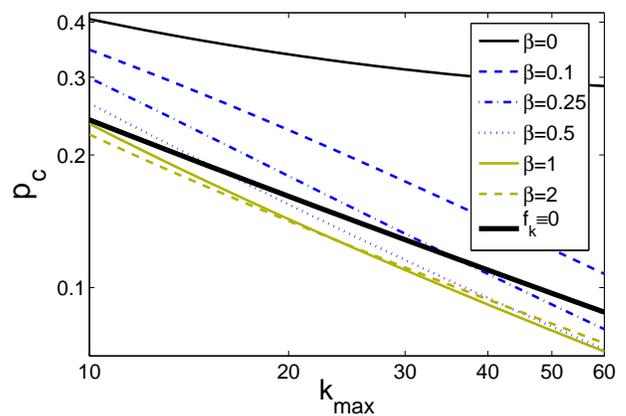}
 \caption{(Color online)
Bond percolation threshold $p_c$ for clustered  networks with
degree distribution $P_k \propto k^{-2.5}$ and cutoff degree
$k_\text{max}$. The fraction $f_k$ of individuals of degree $k$
which are members of households ($k$-cliques) is
$f_k=\lb(2/(k-1)\rb)^{\beta}$ with $\beta$ taking the values from
0 to 2. The thick black curve shows the percolation threshold
$p^\text{rand}_c$ in the unclustered ($f_k\equiv 0$) case.}
\label{f:3}
\end{figure}

\subsection{Examples}
Figure~\ref{f:2} shows the bond percolation threshold $p_c$
calculated from equation~(\ref{GCCcond}) for networks with a
Poisson degree distribution $ P_k=z^k e^{-z}/{k!}$. The log-log
plots show $p_c$ as a function of the mean degree $z=\lb<k\rb>$,
and for clique fractions $f_k$ of the form
\begin{equation}
f_k = \lb(\frac{2}{k-1}\rb)^\beta \quad\text{ for }k\ge 3,
\end{equation}
with $f_k=0$ for $k<3$ (since $k$-cliques only exist for $k\ge
3$). We show  results for values of $\beta$ ranging from 0 (giving
$f_k\equiv 1$ for all relevant $k$) to $\beta=2$ as described in
the caption. Also shown (as a thick black curve) is the
percolation threshold $p_c^\text{rand}=1/z$ in the corresponding
unclustered network. For all values of $\beta$ greater than zero,
we find $p_c>p_c^\text{rand}$ for small values of the mean degree
$z$, but for sufficiently large $z$ the clustered percolation
point $p_c$ becomes slightly less than the configuration model
value $p_c^\text{rand}$. Figure~\ref{f:2}(b) highlights this
clustering-induced decrease of the threshold value by showing that
the ratio $p_c/p_c^\text{rand}$ is (slightly) less than unity for
the larger $z$ values shown.

Figure~\ref{f:3} shows $p_c$ values for the truncated power-law degree distribution
\begin{equation}
P_k = \lb\{\begin{array}{cl} A\, k^{-\gamma} & , 3\le k\le k_\text{max} \\
0  & \text{, otherwise} \end{array}\rb. ,
\label{Pkpowerlaw}
\end{equation}
for $\gamma=2.5$ and with the normalization constant $A$ chosen so
that $\sum P_k = 1$.  The $f_k$ dependence as in Figure~\ref{f:2}.
For convenience we have taken $P_k=0$ for $k\le 2$; this choice
ensures the undamaged graph is relatively well-connected, and in
particular that for larger values of $\gamma$ a GCC exists in the
unclustered network~\cite{Serrano06b}. Note that here the results
are presented as functions of the cutoff degree $k_\text{max}$ in
order to highlight interesting behavior in the
$k_\text{max}\to\infty$ limit of scale-free networks. The results
for the power-law degree distribution are qualitatively similar to
those for the Poisson degree distribution, i.e., in all instances,
except the $\beta=0$ case of constant $f_k$, the clustered
networks show a decrease of $p_c$ with increasing $k_\text{max}$.
At large values of $k_\text{max}$ we see $p_c$ dipping below
$p_c^\text{rand}$ to a greater extent in Figure~\ref{f:3} than for
the Poisson degree distribution in Figure~\ref{f:2}. In the
$\beta=0$ case of constant $f_k$ the clustered threshold $p_c$
always exceeds $p_c^\text{rand}$; some implications of this are
considered in section \ref{sect:SFN} below.

\subsection{Analytical bounds}\label{sect:bounds}
Some insight into these results may be gained by examining explicit bounds for $p_c$ which may be obtained analytically from equation~(\ref{GCCcond}). Since $D_k(p)$ is a monotone function of $p$, by replacing $D_k(p)$ with its respective bounds from~(\ref{ineq}), we can solve~(\ref{GCCcond}) for lower and upper bounds $p_-$ and $p_+$ on the value of $p_c$. Thus we obtain $p_-\le p_c\le p_+$, with
\begin{align}
\nn p_- &= \frac{\lb<k(1-f_k)+f_k\rb>}{\lb<(k-1)(k(1-f_k)+f_k)\rb>}, \\
p_+ &= \frac{\lb<k(1-f_k)+f_k\rb>}{\lb<k(k-1)(1-f_k)\rb>}.
\label{plowupp}
\end{align}
Note that $p_-$ and $p_+$ both reduce to $p_c^\text{rand}$ when $f_k\equiv 0$. We now examine the quantities $p_- - p_c^\text{rand}$ and $p_+ - p_c^\text{rand}$ for some specific
forms of the clique fractions $f_k$. Of particular interest are cases where $p_- - p_c^\text{rand}$ can be shown to be positive, or where $p_+ - p_c^\text{rand}$ is negative. In
the former case we obtain $p_c > p_- > p_c^\text{rand}$, and so can guarantee that the presence of such clustering increases the percolation threshold above $p_c^\text{rand}$; in the latter case we similarly guarantee that $p_c < p_c^\text{rand}$. After a little manipulation, we obtain the expressions
\begin{align}
p_- -p_c^\text{rand} &=
\frac{\lb<k\rb>\lb<k(k-1)f_k\rb> - \lb<k^2\rb>\lb<(k-1)f_k\rb>}{\lb<k(k-1)\rb>\lb<(k-1)(k(1-f_k)+f_k)\rb>},
\label{eqn13a}\\
p_+ -p_c^\text{rand} &= \frac{\lb<k\rb>\lb<(k^2-1)f_k\rb> - \lb<k^2\rb>\lb<(k-1)f_k\rb>}{\lb<k(k-1)\rb>\lb<k(k-1)(1-f_k)\rb>}.
\label{eqn13b}
\end{align}
As the denominators are manifestly positive, the signs of these expressions are determined by the signs of their respective numerators.
\subsubsection{Clustering increases the percolation threshold when $f_k$ is constant}\label{sect:constantF}
We first examine $p_- -p_c^\text{rand}$ in the case where $f_k=F$, a constant, for all $k\ge 3$. The numerator of~(\ref{eqn13a}) then simplifies to
\begin{equation}
F\lb( \lb<k^2\rb>-\lb<k\rb>^2+\lb<k^2\rb>(P_2-P_0) - 2\lb<k\rb> P_2\rb).
\label{lowb}
\end{equation}
For the power-law degree distribution~(\ref{Pkpowerlaw}) we have $P_k=0$ for $k<3$, and so this expression reduces to $F\, \text{var}(k)$ where $\text{var}(k)$ is the variance $\lb<k^2\rb>-\lb<k\rb>^2$ of the degree distribution. Since this is positive for any $k_\text{max}>3$, we have proven that $p_c > p_c^\text{rand}$ for constant $f_k$ in this case. Similarly, it can be shown that~(\ref{lowb}) is positive, and hence  $p_c > p_c^\text{rand}$, for the Poisson degree distribution. These results are consistent with the $\beta=0$ results for $p_c$ (thin black lines) in Figures~\ref{f:2} and~\ref{f:3}, which never dip below the $p_c^\text{rand}$ values (thick black line).

\subsubsection{Clustering decreases the percolation threshold if $f_k=F/(k-1)$}\label{sect:decreasingF}
Next, we consider the numerator of $p_+ - p_c^\text{rand}$ for $f_k$ of the form $F/(k-1)$ for $k\ge 3$, with $F$ in the range $0<F\le 2$. The numerator of (\ref{eqn13b}) then simplifies to
\begin{align}
\nn F\Bigr( \lb<k\rb>^2 + \lb<k\rb> - \lb<k^2\rb> - \lb<k\rb>& (P_0+2P_1+3P_2)+ \\
&\lb<k^2\rb> (P_0+P_1+P_2)\Bigr).
\end{align}
For the power-law degree distribution~(\ref{Pkpowerlaw}) this further reduces to $\lb<k\rb> - \text{var}(k)$, and as $k_\text{max}\to \infty$ this certainly becomes negative. Specifically, for the exponent $\gamma=2.5$ used in Figure~\ref{f:3}, this bound guarantees that $p_c$ is less than $p_c^\text{rand}$ for $k_\text{max}\ge 13$. This is consistent with the curve for $\beta=1$ in Figure~\ref{f:3}. For the Poisson distribution, the numerator simplifies to $F z^4 e^{-z}/2$---however, as this quantity is positive we cannot draw any strong conclusions for this case.

\subsection{Scale-free networks}
\label{sect:SFN} A scale-free network (SFN) has degree
distribution $P_k\propto k^{-\gamma}$ with $2<\gamma<3$ for
sufficiently large $k$. Networks with such degree distributions
may be generated by taking the limit $k_\text{max}\to \infty$ of
the truncated power-law networks introduced in equation
(\ref{Pkpowerlaw}). Of particular interest is the bond percolation
threshold $p_c$ which is
known~\cite{Cohen00,Callaway00,Pastor-Satorras01,Albert00,Boguna02,Boguna03b,Goltsev08}
to be zero  for randomly wired (uncorrelated) SFNs. This can be
seen from equation~(\ref{allord}): the second moment $\sum k^2
P_k$ for SFNs is infinite while the mean degree $z$ is finite, and
so the denominator of the expression for $p_c^\text{rand}$ grows
without bound as the cutoff $k_\text{max}$ is increased, giving
the result $p_c^\text{rand}\to 0$ as $k_\text{max} \to \infty$.
The results of~\cite{Vazquez03,Boguna03b} indicate that correlated
(assortative) tree-like networks with scale-free degree
distributions also have vanishing percolation threshold,
and~\cite{Serrano06b} and~\cite{Newman03b} hypothesize that
clustering cannot cause the percolation  threshold to be non-zero.
However, Trapman~\cite{Trapman07} has applied his clustering model
to note that if $f_k=1$ for all $k$ then  a non-zero bond
percolation threshold is established even in scale-free networks.
To see this result, it is convenient to express the lower bound
$p_-$ for the percolation threshold given in
equation~(\ref{plowupp}) in terms of the degree distribution
$\wP_k$ of the super-graph, using equation~(\ref{pp}):
\begin{equation}
p_- = \frac{\sum k \wP_k}{\sum k(k-1)\wP_k}. \label{plow2}
\end{equation}
This implies that the lower bound $p_-$ for the percolation
threshold in the individuals graph is equal to the percolation
threshold in the randomly-wired super-graph. In other words, the
individuals graph can only possess a GCC if the super-graph also
has a GCC. Now consider Trapman's example of a SFN where all $f_k$
are equal to one, with degree distribution (\ref{Pkpowerlaw}) and
in the limit $k_\text{max}\to \infty$. The super-graph degree
distribution is then $\wP_k \propto P_k/k \propto k^{-\gamma-1}$.
This degree distribution has finite variance for $\gamma>2$, and
so it follows that the right hand side of~(\ref{plow2}) is
non-zero. In fact, we can explicitly evaluate $p_-$ to obtain the
following bound on the percolation threshold:
\begin{equation}
p_c \ge \frac{\sum k^{-\gamma}}{\sum (k-1)k^{-\gamma}} =
\frac{1}{z-1}, \label{SFNbound}
\end{equation}
where $z$ is the (finite) mean degree of the individuals
scale-free network, $z=\sum k P_k$.

It is worth pointing out that the mechanism described here for
generating a non-zero percolation threshold in SFNs is distinctly
different from those previously examined for tree-like correlated
networks~\cite{Vazquez03}, 2D lattice-embedded
networks~\cite{Warren02}, and for clustered growing
networks~\cite{Eguiluz02}. All of these examples are
disassortative networks, i.e., the average degree of neighbors of
$k$-degree nodes $\lb<k\rb>_{\text{nn}}$ is a decreasing function
of $k$ (with an asymptotic constant value as $k\to\infty$ in the
case of~\cite{Eguiluz02}). By contrast, the individuals graph
generated by Trapman's model with $f_k\equiv F=1$ is strongly
assortative, since high-degree nodes link almost exclusively to
nodes of the same degree.  Indeed, we show in Appendix B that the
joint pdf $P(k,j)$ of degrees of vertices at either end of a
randomly chosen edge in the individuals graph is
\begin{equation}
P(k,j) = \frac{P_j}{z}\lb(P_k+(j-1)\delta_{k j}\rb). \label{Pkj}
\end{equation}
Hence the average degree of neighbors of nodes with degree $k$ is
$\lb<k\rb>_{\text{nn}}=k-1+\frac{z}{k}$
and so increases linearly for large $k$.

We also highlight the fact that the non-zero percolation threshold
arising in the $F=1$ Trapman model is due to the clustering, and
not just a result
  of the degree-correlations induced by the clique structure. Indeed, consider a
  correlated but unclustered (tree-like) network with degree-degree correlations equal to those given by~(\ref{Pkj}),
   and reintroduce the cutoff $k_\text{max}$ for the SFN degree distribution. The
   percolation threshold for such unclustered networks is known~\cite{Vazquez03,Boguna03b} to be
    given by the reciprocal of the largest eigenvalue of the matrix $C$ with entries
     $C_{k j} = (j-1)z P(k,j)/(k P_k)$. In the present case this  threshold scales as
     $k_\text{max}^{-1}$ as $k_\text{max} \to \infty$, and so the correlated
     tree-like network
    has a vanishing percolation threshold. This is consistent with the behavior of strongly assortative
    tree-like networks studied in~\cite{Vazquez03}, and shows that
    the finite threshold given by (\ref{SFNbound}) is directly
    attributable to the non-zero clustering in the Trapman model. Criteria for the existence of a
    finite SFN percolation threshold for non-constant $f_k$ will be
    reported elsewhere.

\subsection{Real-world networks}
\begin{table} [floatfix]
\centering
\begin{tabular}{|l|c|c|c|c|c|}
  \hline
Network                             &$p_-$  & $p_+$  & $p_c$  & $p_c^\text{rand}$ \\
  \hline
  Power Grid~\cite{Watts98, power_url}                  & 0.3580 & 0.3739 & 0.3645 & 0.3483 \\
  AS Internet~\cite{asrel_url}                      & 0.0031 & 0.0031 & 0.0031 & 0.0035 \\
  Collaborations~\cite{Newman01b, cond_mat_url}             & 0.0273 & 0.0279 & 0.0279 & 0.0380 \\
  World Wide Web~\cite{Albert99, WWW_url}               & 0.0020 & 0.0020 & 0.0020 & 0.0036 \\
  Router-Level Internet~\cite{ITDK_url}                 & 0.0244 & 0.0245 & 0.0245 & 0.0271 \\
  PGP Network~\cite{Guardiola02, Boguna04, PGP_url}         & 0.0545 & 0.0567 & 0.0561 & 0.0559 \\
  \hline
\end{tabular}
\caption{(Color online)  Values of the bounds $p_-$ and $p_+$
(using equation~(\ref{plowupp})), the bond percolation threshold
in the clustered model network $p_c$ (using
equation~(\ref{GCCcond})), and the randomly-wired percolation
threshold $p_c^\text{rand}$ (using equation~(\ref{allord})) for
some real-world networks. The bounds are calculated using the
degree distribution $P_k$ and clustering $c_k$ (for $k>2$) of the
real-world network, converting clustering to $k$-clique fractions
via equation~(\ref{cf2}).
} \label{table1}
\end{table}
In Table~\ref{table1} we show the results of applying  our model
of clustering to some real-world networks. Given the degree
distribution $P_k$ and the degree-dependent clustering $c_k$ of a
real-world network, we choose $f_k$ values using
equation~(\ref{cf}) so that the model network has a $k$-clique
structure which matches to $P_k$ and (for all $k\ge3$, and
provided $c_k$ is not too large) to $c_k$:
\begin{equation}
f_k =\text{min}\lb(1,\frac{ k \,c_k }{k-2} \rb)\quad\text{ for
}k\ge 3. \label{cf2}
\end{equation}
Using equations~(\ref{plowupp}) and~(\ref{allord}) we calculate
the bounds $p_-$ and $p_+$ for the percolation threshold in the
clustered model network, as well as the threshold
$p_c^\text{rand}$ for the corresponding randomly-wired graph. In
most cases (the PGP network being the exception) we can
immediately see from the bounding values $p_-$ and $p_+$ whether
the clustered percolation threshold $p_c$ will exceed
$p_c^\text{rand}$ or not. For the power grid network we have
$p_->p_c^\text{rand}$ and so conclude that clustering increases
the percolation threshold. For PGP the bounds are inconclusive,
but calculation using equation~(\ref{GCCcond}) confirms
$p_c>p_c^\text{rand}$ in this case also. For all other networks
studied we find $p_+<p_c^\text{rand}$, so that clustering
decreases the percolation threshold.

We obtain these results on $p_c$ under the assumption that the
generalized Trapman model can describe the structure of real-world
networks by matching the degree distribution and degree-dependent
clustering. This is admittedly a rather strong assumption, and
further verification is needed before these results can be
considered more just some interesting examples of applying the
model. As percolation thresholds are defined only in the  infinite
system size limit $N\to\infty$, it is not possible to directly
calculate percolation thresholds for (necessarily finite)
real-world networks, but it appears from Figure 7
of~\cite{Serrano06b} that the PGP network percolates for $p<0.05$,
whereas the $p_c$ value we predict in Table~\ref{table1} is
substantially larger. We conclude that the Trapman model is not
necessarily a good predictor of the percolation properties of
real-world networks, despite its ability to match the degree
distribution and the clustering of the network.

In summary, in this section we have derived  the polynomial
equation~(\ref{GCCcond}) for the percolation threshold $p_c$ in
the presence of clustering, and solved it numerically for some
examples. Analytical bounds on the value of $p_c$ have also been
derived, and for the truncated power-law degree distribution
clique fractions of the form $f_k=F$ and $f_k=F/(k-1)$ have been
respectively shown to guarantee that $p_c$ is greater than, or
less than, the unclustered threshold value $p_c^\text{rand}$.
Application of the model to some real-world networks yields
examples where $p_c<p_c^\text{rand}$ in some cases, with
$p_c>p_c^\text{rand}$ in others. In scale-free
  networks clustering with $f_k\equiv 1$ guarantees a finite
  percolation threshold, in contrast to corresponding tree-like networks (even those with
  the same degree correlations) where the percolation threshold
  vanishes.

\section{Calculating GCC sizes} \label{s:GCC}
In this section we develop an analytical approach  to calculating
the size of the giant connected component in the damaged
individuals graph with bond occupation probability $p$. In an
epidemiological context, the GCC size corresponds to the expected
size of epidemic outbreaks in the population. Of particular
interest is the effect of clustering on the epidemic size.

Our method is based on a general  formulation for cascade sizes on
random networks, described in detail in~\cite{Gleeson08a}. We note
that a generating function approach could also be used here,
similar to~\cite{Newman03b}, and such a method could yield the
full distribution of connected component sizes. However our method
has the advantage of being  readily generalizable  to the study of
other cascade-type problems on networks, as we show in
section~\ref{s:kcore} by using it to calculate the size of
$k$-cores in the clustered networks. The method is a
generalization of the approach of Dhar et al.  for the
zero-temperature random-field Ising model on a network
\cite{Dhar97} and has been successfully applied to cascade
dynamics in various models \cite{Gleeson07a,Gleeson08b}, including
the calculation of $k$-core sizes in correlated (but unclustered)
networks \cite{Gleeson08a}.

\begin{figure}[ht!]
\centering
\includegraphics[width=0.6\columnwidth]{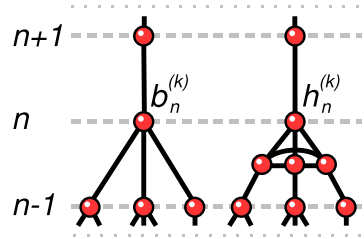}
\caption{(Color online) Schematic showing  parts of a tree
approximation for a super-individuals graph which is expanded to
an individuals graph. Level $n$ is occupied by a bachelor (left)
and by a top node of the expanded household (right). Other members
of the same household are located at an intermediate level. }
\label{fig_tree}
\end{figure}
Following the approach of~\cite{Gleeson08a}, we approximate the
randomly wired super-graph as a tree structure. This tree ansatz
is commonly used  for the configuration model; it assumes the
absence of finite loops in the super-graph in the $\widetilde
N\to\infty$ limit and allows only  the infinite loops whose
presence permits the use of mean-field theory
\cite{Dorogovtsev08}. Figure~\ref{fig_tree} shows part of such a
structure, with the super-individuals  now expanded to show the
individual nodes which constitute households. We label the levels
of the tree as shown, with each super-individual at level $n$
having a single parent at level $n+1$. Degree-$k$ bachelors at
level $n$ therefore have $k-1$ children at level $n-1$; degree-$k$
households at level $n$ are considered to consist of a \emph{top
individual} (shown at level $n$), with the $k-1$ other individuals
of the household drawn at an intermediate level. Each of these
$k-1$ individuals has one child super-individual at level $n-1$.

The cascade-based approach to calculating the expected  size of
the giant connected component is as follows. Having chosen a value
for the bond occupation probability $p$ we damage the individuals
graph by deleting each link  between individuals with probability
$1-p$. We label nodes which are part of a connected component of
the graph as \emph{active}, with the remaining nodes termed
\emph{inactive}. A random individual is selected as the top (i.e.
root) of a tree, with his first neighbors on the next lower level,
their neighbors at the following lower level, and so on. To
determine the steady-state fraction of active nodes in the
network, we must determine the probability that the individual at
the top of the tree is active. Note all nodes in the tree are
initially inactive, and that once a node is activated it cannot
later become inactive.  Starting at level 0 (the bottom of the
tree), we examine the propagation of activity from level $n$ to
level $n+1$, proceeding one level at a time and using the fact
that nodes at level $n+1$ are inactive until their children cause
them to become active.

Define $q_n$ as the
probability that a super-individual at level $n$ is active
\footnote{Because all nodes are initially inactive in the cases
studied here, we do not require this probability to be conditional
on the inactive state of the parent, as used in \cite{Dhar97,
Gleeson07a, Gleeson08a}.}.
 Similar 
 probabilities may be defined
separately for households and for bachelors; moreover we
distinguish between super-individuals of different degree $k$.
Denote by $b_n^{(k)}$ the probability that a bachelor node of
degree $k$ at level $n$ is active,
and by $h_n^{(k)}$ the probability that the top individual node in
a household of degree $k$ is active.
Since a randomly-chosen super-individual connects to a
super-individual of degree $k$ with probability $k \wP_k/\wz$, we
have the relation
\begin{equation}
q_{n+1} = \sum_{k=1}^\infty \frac{k}{\wz} \wP_k\lb((1-g_k)b_{n+1}^{(k)}+g_k h_{n+1}^{(k)}\rb).
\label{qupdate}
\end{equation}

To determine $b_{n+1}^{(k)}$ and $h_{n+1}^{(k)}$ in terms of $q_n$
we consider how the  property of being active (i.e. being a member
of a connected component) propagates from level to level.
As we move focus from
level to level, we need only consider the active fraction at level
$n$ to determine how many nodes at level $n+1$ change from their
inactive initial state. For bachelor nodes of degree $k$, we need
consider only their $k-1$ children at level $n$. Each of the
children is part of a connected component with probability $q_n$
and the link to this child is undamaged with probability $p$. The
bachelor node becomes active if any one of the $k-1$ links to
level $n$ yield an undamaged connection to an active child, thus
we have the update rule~\cite{Gleeson08a}
\begin{equation}
b_{n+1}^{(k)} = 1-(1-p\, q_n)^{k-1}.
\label{bupdate}
\end{equation}
For households at level $n+1$ we consider the situation of the top individual. Within the $k$ individual nodes of the household, the top individual is part of a connected cluster of $m$ individuals with probability $P(m|k)$ (see Appendix A). Each of the $m-1$ other individuals within the household has one edge linking to level $n$, and so the probability that at least one of these will become active is $1-(1-p\, q_n)^{m-1}$. Summing over the possible values of $m$, we obtain the probability of the top node of the household becoming active:
\begin{equation}
h_{n+1}^{(k)} = \sum_{m=1}^k P(m|k) \lb(1-(1-p\,q_n)^{m-1}\rb).
\label{hupdate}
\end{equation}
Combining~(\ref{qupdate}),~(\ref{bupdate}) and~(\ref{hupdate}) enables us to write a single update equation for $q_n$ of the form $q_{n+1} = G(q_n)$ with
\begin{align}
G(q) = \lb. \sum_{k=1}^\infty \frac{k}{\wz} \wP_k \rb[ &(1- g_k) \lb(1-(1-p\, q)^{k-1}\rb) \label{Geqn} \\
\nn &+ \lb. g_k \sum_{m=1}^k P(m|k)\lb( 1-(1- p\, q)^{m-1}\rb) \rb].
\end{align}
Starting from an infinitesimally small positive value (e.g.,
$q_0=1/N$ as $N\to \infty$), this equation is iterated to yield
the steady-state solution $q_\infty$ corresponding to an infinite
network. Finally, we consider the individual at the top (or root)
of this infinite tree. Suppose the individual has degree $k$ (this
happens with probability $P_k$) and so has $k$ children. With
probability $1-f_k$ it is an individual who was a bachelor in the
super-graph, and so is activated by its children with probability
$1-(1-p\, q_\infty)^k$. Otherwise it is a member of a household of
size $k$, and so is part of a connected cluster of $m$ individuals
within this household with probability $P(m|k)$. The whole cluster
becomes active is any member of it has an undamaged link to an
active child; this happens with probability $1-(1-p\,
q_\infty)^m$. Putting together all the possibilities, we obtain an
expression for $S$, the expected size of the giant connected
component:
\begin{align}
\nn S = \lb.\sum_{k=0}^\infty P_k \rb[ &(1-f_k) \lb(1-(1-p\,q_\infty)^k\rb) \\
& \lb. + f_k\sum_{m=1}^k P(m|k)\lb( 1- (1- p\,q_\infty)^m\rb) \rb]
\label{Seqn},
\end{align}
where $q_\infty$ is the steady-state of the  iteration
$q_{n+1}=G(q_n)$ defined by equation~(\ref{Geqn}). Indeed, the
iteration process with infinitesimal $q_0$ can be seen as a
solution method for the self-consistent equation $q_\infty =
G(q_\infty)$. 

Classical results on uncorrelated, unclustered networks are
recovered by setting $f_k=g_k=0$ for all $k$ in equations
(\ref{Geqn}) and (\ref{Seqn}); this reduction (via the notation
mapping $1-p q \mapsto x$) recovers, for example, equations (9)
and (14) of \cite{Dorogovtsev08}.

Note that a general cascade condition~\cite{Gleeson08a} for this system requires
\begin{equation}
\frac{d G}{d q} > 1 \quad \text{ at } q=0,
\end{equation}
in order that the initial iterations of the relation
$q_{n+1}=G(q_n)$ allow $q_n$  to grow finitely large. The lowest
value of $p$ for which this condition holds defines the bond
percolation threshold $p_c$, and it is easy to check that this
condition reduces to equation (\ref{orgcond}), which was derived
using more traditional arguments in section~\ref{s:pc}.

\begin{figure}
\centering
\includegraphics[width=0.95\columnwidth]{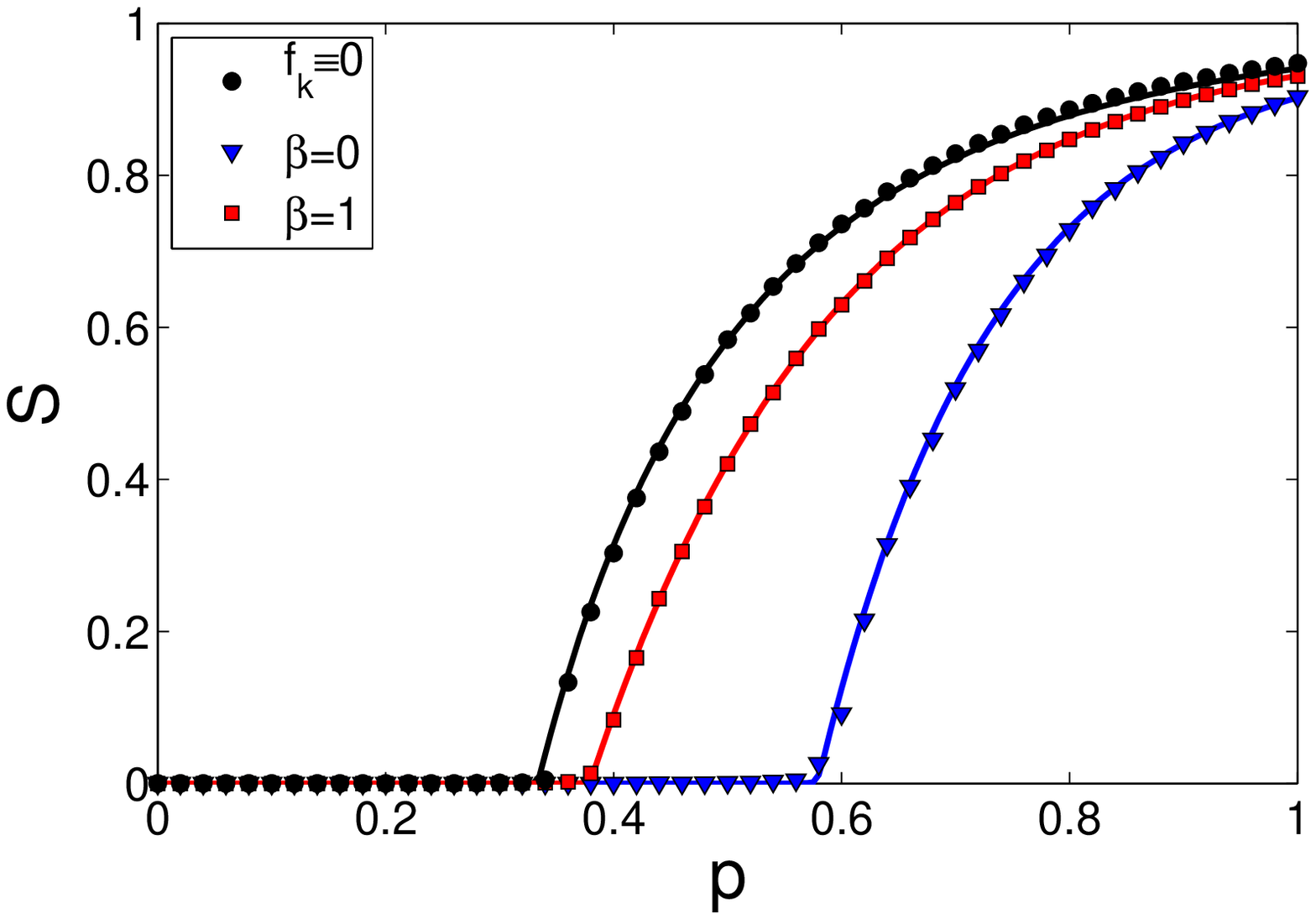}
\put(-240,140){\bf(a)}
\hspace{0.3cm}
\includegraphics[width=0.95\columnwidth]{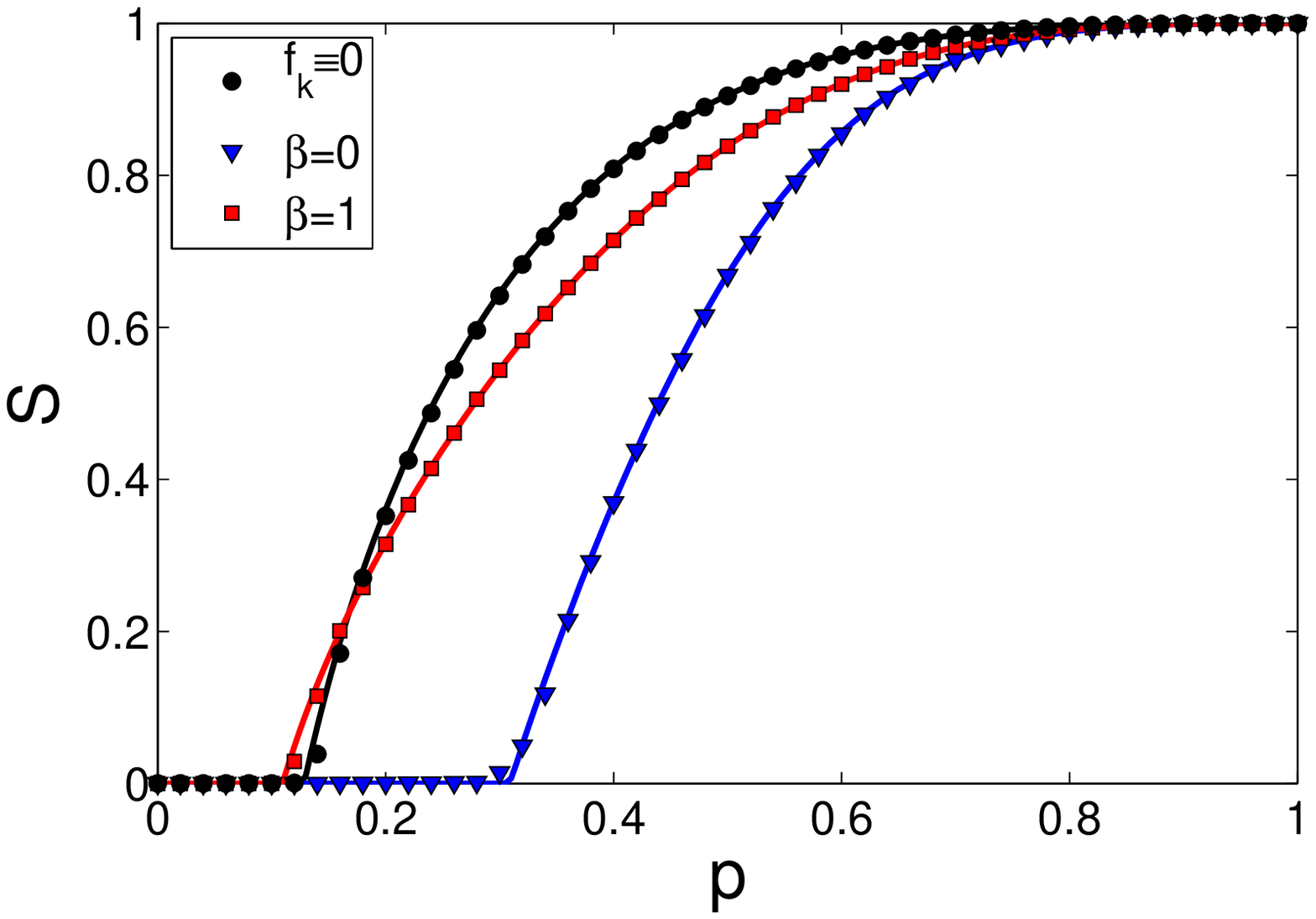}
\put(-240,140){\bf(b)} \caption{(Color online)  Size of giant
connected component $S$ as a function of bond occupation
probability in (a) clustered Poisson random graphs with mean
degree $z=3$, and in (b) clustered graphs with truncated power-law
degree distribution $P_k\propto k^{-2.5}$ for $3\le k\le
k_\text{max}$, with $k_\text{max}=30$ here. Symbols are the
results of numerical simulations on a single network with $N=10^5$
individuals (averaged over 10 realizations of the percolation
process), and curves show the analytical result from equations
(\ref{qupdate}) and (\ref{Seqn}). The fraction $f_k$ of
individuals of degree $k$ which are members of households
($k$-cliques) is $f_k=\lb(2/(k-1)\rb)^{\beta}$ with $\beta$ taking
values indicated in the legend. The unclustered case ($f_k\equiv
0$) is also shown for comparison. } \label{f:5}
\end{figure}
Figures~\ref{f:5}(a) and~\ref{f:5}(b) show a comparison  between
the analytical solution (curves) and numerical computation of GCC
sizes in networks generated using the algorithm of section
\ref{s:GenClustNet} with $N=10^5$ individuals
(symbols)~\footnote{The number $\widetilde N$ of super-individuals
in step (i) of the algorithm of section~\ref{s:GenClustNet} was
tuned to give $N=10^5$ nodes in the individuals graph.}. The
degree distributions of the networks of Figure~\ref{f:5}(a) are
Poisson (as in Figure~\ref{f:2}) with mean degree $z=3$, while the
networks for Figure~\ref{f:5}(b) have a truncated power-law degree
distribution (\ref{Pkpowerlaw}) with $k_\text{max}=30$ (cf.
Figure~\ref{f:3}). Note that the values of the percolation
threshold $p_c$ predicted in Figures~\ref{f:2} and~\ref{f:3}
correspond to the $p$ values where the GCC size becomes non-zero.
For the Poisson case both cases with clustering have $p_c$ larger
than the unclustered value $p_c^\text{rand}$, while the power-law
case of Figure~\ref{f:5}(b) shows that $p_c$ may be larger or
smaller than the unclustered value, depending on the form of the
$k$-clique fraction $f_k$. The agreement between theory and
numerical results is excellent.

\section{Calculating k-core sizes} \label{s:kcore}

The $k$-core
of a network is the largest subgraph whose nodes have degree at least $k$. As discussed in~\cite{Dorogovtsev06, Goltsev06, Gleeson08a}, the size of the $k$-core may be calculated as the steady state of a cascade process. We consider the nodes of the individuals network to have two  possible states, labelled \emph{pruned} and  \emph{unpruned}, and begin with all nodes in the unpruned state. In the first step of the cascade process for calculating the $k$-core for $k=K$, all nodes with fewer than $K$ neighbors are relabelled as pruned---these nodes cannot be part of the the $K$-core. In each subsequent iteration, any node with fewer than $K$ unpruned neighbors is relabelled as pruned. In other words, a node of degree $k$ becomes pruned if the number $k-m$ of its unpruned neighbors is smaller than $K$. In the steady-state limit of this cascade process, precisely those nodes in the $K$-core remain unpruned. 

The cascade-based approach of section~\ref{s:GCC} can be applied
to  calculate $k$-core sizes in the clustered networks generated
by the generalized Trapman model of section~\ref{s:GenClustNet}.
Similar to the discussion preceding equation~(\ref{qupdate}), we
begin with the creation of a tree whose top (or root) is a
randomly selected node of the network. All nodes in the tree are
initially in the unpruned state, and we examine the propagation of
the pruned fraction from level $n$ to level $n+1$ in the tree,
proceeding one level at a time. Our goal is the determination of
the probability that the top (or root) of the tree is pruned; this
gives the final fraction of pruned nodes in the original network.
 We
define $q_n$ as the 
 probability that a
super-individual at level $n$ is pruned.
Similarly, denote by $b_n^{(k)}$ the probability that a bachelor
node of degree $k$ at level $n$ is pruned,
 and by $h_n^{(k)}$ the
probability that the top individual node in a household of degree
$k$ is pruned.
Equation~(\ref{qupdate}) of section~\ref{s:GCC} then applies
directly, and it remains only to define the updating rules for
$b_{n+1}^{(k)}$ and $h_{n+1}^{(k)}$.

To this end it is convenient to introduce response functions $F_b(m,k)$ and $F_h(m,k)$ which respectively denote the probabilities that a $k$-degree bachelor or a $k$-degree household become pruned when they have $m$ pruned neighbors. A bachelor becomes pruned when it has less than $K$ unpruned neighbors, i.e. when
$k-m<K$; otherwise it remains unpruned. Therefore the bachelor response function is given by (see equation (10) of \cite{Gleeson08a})
\begin{equation}
F_b(m,k) =\lb\{\begin{array}{cl} 1 & , k-m<K  \\
0 & , k-m\ge K \end{array}\rb. \label{Fb}.
\end{equation}

For households we must take account of the $k$-clique structure.
First, if $k<K$ then every node in the $k$-clique has less than
$K$ (unpruned) neighbors, and so the entire household is
immediately pruned. Also, for the case $k=K$, the whole household
becomes pruned if any one of its neighbors is  pruned, i.e. if
$m>0$, and remains unpruned otherwise. Finally, no node in a
$k$-clique can become pruned if $k>K$, because in this case an
individual of degree $k$ needs at least two pruned neighbors in
order to become pruned itself, but each node in the $k$-clique has
only one external neighbor (and all nodes in the $k$-clique are
initially unpruned). Thus it is straightforward to see that
\begin{equation}
F_h(m,k) =\lb\{\begin{array}{cl}
    F_b(m,k) &, k\le K \\
    0 & , k>K
\end{array}\rb.
\label{Fh}.
\end{equation}

Next, since each child at level $n$ is independently pruned with
probability $q_n$, a bachelor or household of degree $k$ has
exactly $m$ out of $k-1$ children pruned with probability
$\binom{k-1}{m} q_n^m(1-q_n)^{k-1-m}$. Therefore, summing over
every possible number of pruned children $m$ gives the probability
that a bachelor node of degree $k$ at level $n$ is  pruned:
\begin{equation}
b_{n+1}^{(k)} = \sum\limits _{m=0}^{k-1}\binom{k-1}{m} q_n^m(1-q_n)^{k-1-m}F_b(m,k),
\label{bK_F}
\end{equation}
and a similar expression for a household can be written using equation~\eqref{Fh} as
\begin{equation}
h_{n+1}^{(k)} =\lb\{\begin{array}{cl}
    b_{n+1}^{(k)} &, k\le K \\
    0 & , k>K
\end{array}\rb.
\label{hK_F}.
\end{equation}

Note that $b_{n+1}^{(k)}$ and $h_{n+1}^{(k)}$ can also be written in a less general form without the use of response functions as
\begin{align}
b_{n+1}^{(k)} &= \lb\{ \begin{array}{cl}
    1 & , k<K \\
    \sum\limits _{m=k-K+1}^{k-1}\binom{k-1}{m} q_n^m(1-q_n)^{k-1-m} & , k\ge K
    \end{array}\rb.,\label{bK}\\
\nn \text{and}\\
h_{n+1}^{(k)} &= \lb\{\begin{array}{cl} 1 & , k<K \\
1-(1-q_n)^{k-1} & , k= K\\
0 &, k>K\end{array}\rb. .\label{hK}
\end{align}
Using the update rules~\eqref{bK_F} and~\eqref{hK_F} (or alternatively~\eqref{bK} and~\eqref{hK}) in conjunction with~(\ref{qupdate}) enables us to iterate from an infinitesimally small positive $q_0$ to the steady state $q_\infty$ corresponding to an infinite network.

Finally, consider the individual at the top (or root) of the infinite tree, assuming it has degree $k$, i.e., $k$ children. With probability $1-f_k$ it was a bachelor in the super-graph, and by analogy with~\eqref{bK_F} is pruned with probability
\begin{equation}
\rho_b^{(k)} = \sum\limits_{m=0}^{k} \binom{k}{m} q_\infty^m(1-q_\infty)^{k-m}F_b(m,k).
\label{brho_F}
\end{equation}
Similarly, if the individual is in a household (with probability $f_k$), it is pruned with probability
\begin{equation}
\rho_{h}^{(k)} =\lb\{\begin{array}{cl}
    \rho_b^{(k)} &, k\le K \\
    0 & , k>K
\end{array}\rb. .
\end{equation}
Then the final density of pruned nodes in the individuals network is given by (cf. equation (\ref{Seqn}))
\begin{equation}
\rho = \sum_{k=0}^\infty P_k \lb[ (1-f_k)\rho_b^{(k)}+f_k \rho_h^{(k)}\rb],
\label{rhoeqn}
\end{equation}
and the fractional size of the $k$-core for $k=K$ is given by $1-\rho$.

We can combine equations (\ref{qupdate}), (\ref{bK_F}) and (\ref{hK_F}) to give an explicit self-consistent equation for $q_\infty$:
\begin{align}
q_\infty &= H(q_\infty) \nonumber\\
&\equiv \sum_{k=1}^\infty \frac{k}{\widetilde z}{\widetilde P}_k
\sum_{m=0}^{k-1} \binom{k-1}{m} q_\infty^m (1-q_\infty)^{k-1-m}
W_k F_b(m,k),\nonumber\\ \label{SCK}
\end{align}
where
\begin{equation}
W_k =\lb\{\begin{array}{cl} 1 & , k \le K \\
1-g_k & , k> K\end{array}\rb. \label{Wk}.
\end{equation}
The iteration process for $q_n$ starting from infinitesimal $q_0$ converges to the lowest solution of the self-consistent equation (\ref{SCK}).

The analysis of section IV of \cite{Goltsev06} may be applied here
to provide an interpretation for $q_\infty$ in terms of measurable
quantities on the network. Let $L_K$ be the number of edges in the
super-graph which connect two individuals belonging to the
$K$-core, and let $L$ be the total number of edges in the
super-graph. Then, as shown in Appendix C,
\begin{equation}
(1-q_\infty)^2 = \frac{L_K}{L}, \label{eqn35}
\end{equation}
i.e., the quantity $q_\infty$ is related to the fraction of
super-graph edges which link individuals in the $K$-core.

 In the limit of zero clustering ($f_k=g_k=0$ for all
$k$), equations   (\ref{rhoeqn}) and (\ref{SCK}) reduce to
existing results for $k$-cores on (undamaged) configuration model
networks, as in equations (1) and (2) of \cite{Dorogovtsev06} via
the mapping of notation $q\mapsto R$, $\rho\mapsto 1-M$, see
Appendix D.

\begin{figure} [ht!]
\centering
\includegraphics[width=0.95\columnwidth]{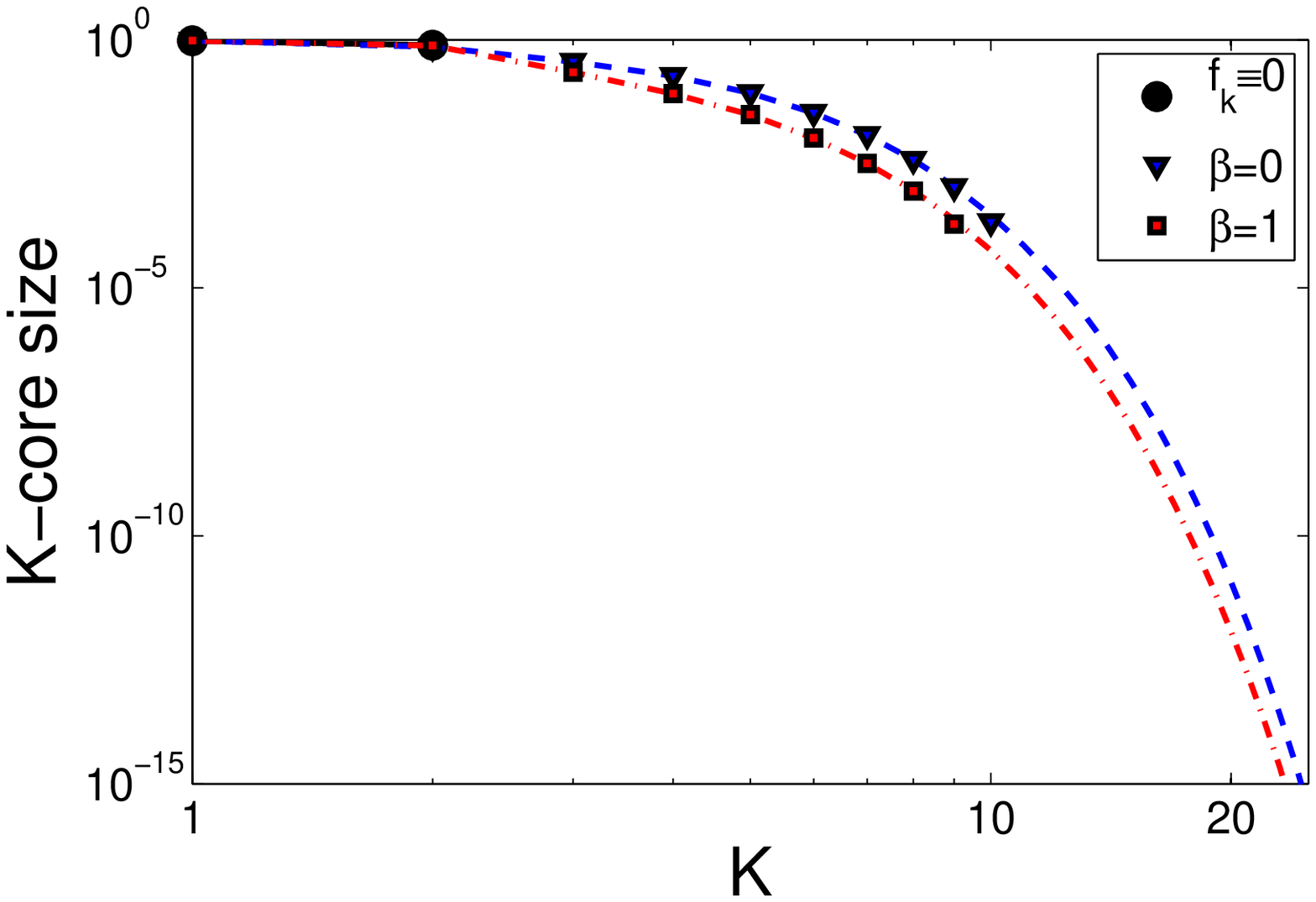}
\put(-240,140){\bf(a)}
\hspace{0.3cm}
\includegraphics[width=0.95\columnwidth]{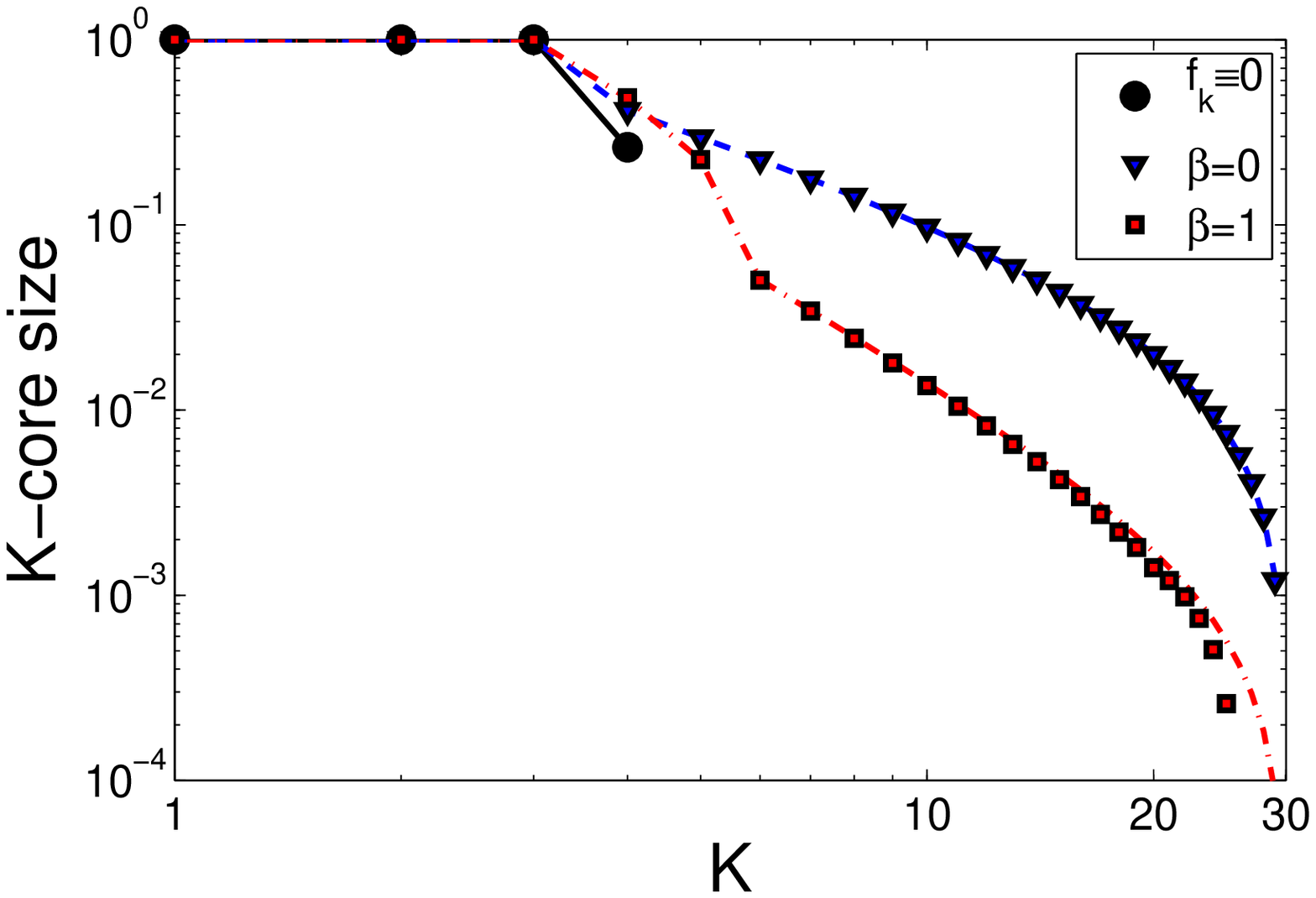}
\put(-240,140){\bf(b)}
 \caption{(Color online)
K-core sizes in (a) clustered Poisson random graph with mean
degree $z=3$, and in (b) clustered graphs with truncated power-law
degree distribution $P_k\propto k^{-2.5}$ for $3\le k\le
k_\text{max}$, with $k_\text{max}=30$ here. Symbols are the
results of numerical simulations on a single network with $N=10^5$
individuals, and curves show the analytical result from
Eq.~\eqref{rhoeqn}. The fraction $f_k$ of individuals of degree
$k$ which are members of households ($k$-cliques) is
$f_k=\lb(2/(k-1)\rb)^{\beta}$ with $\beta$ taking values indicated
in the legend. The unclustered case ($f_k\equiv 0$) is also shown
for comparison.} \label{f:6}
\end{figure}
Figures~\ref{f:6}(a) and~\ref{f:6}(b) show comparisons  between
the theory and numerical calculations of $k$-core sizes on
clustered networks generated by the algorithm of
section~\ref{s:GenClustNet}. Figure~\ref{f:6}(a) is for a network
with Poisson degree distribution with $z=3$ (cf.
Figure~\ref{f:5}(a)). The unclustered ($f_k=0$) case has no
$k$-cores for $k>2$, but the presence of cliques leads to non-zero
$k$-core sizes for all $k$ with $f_k>0$. Since we use finite-size
graphs we cannot numerically resolve $k$-cores of fractional size
smaller than $1/N$ ($N=10^5$ here) but the agreement between
theory and simulation is excellent for $K$ up to approximately 10.
A network with truncated power-law degree
distribution~(\ref{Pkpowerlaw}) with $k_\text{max}=30$ (cf.
Figure~\ref{f:5}(b)) has $k$-core sizes as shown in
Figure~\ref{f:6}(b). Again, non-zero clustering leads to non-zero
$k$-core sizes for all $K$ up to $k_\text{max}$, and agreement
between theory and numerics is excellent except for finite size
effects upon very small $k$-cores.

\section{Conclusions}\label{s:Concl}
We have shown that a generalization of the Trapman
model~\cite{Trapman06, Trapman07} of clustered clique-tree
networks has several analytically tractable features. These
include the ability to calculate the bond percolation threshold,
size of the giant connected component, and sizes of $k$-cores. The
algorithm for generating realizations of model networks is
described in section~\ref{s:GenClustNet}. The degree distribution
$P_k$ of the network is specified, along with the fraction $f_k$
of $k$-degree nodes residing in $k$-cliques. The parameters $f_k$
are related to the degree-dependent clustering coefficients $c_k$
by equation~(\ref{cf}), and so allow us to tune the level of
clustering in the network.

The main analytical results are equation~(\ref{GCCcond}) for the
 bond percolation threshold, and the iteration schemes of sections~\ref{s:GCC}
  and~\ref{s:kcore} (see equations~(\ref{Seqn}) and~(\ref{rhoeqn})) for the sizes
   of the giant connected component and $k$-cores, respectively. The percolation
    threshold $p_c$ is determined by solving the polynomial equation~(\ref{GCCcond}), see Figures~\ref{f:2} and~\ref{f:3} for examples.
    We have also examined explicit upper and lower bounds for $p_c$
     (see section~\ref{sect:bounds}). Of particular interest is the relationship
      between $p_c$ and the percolation threshold $p_c^\text{rand}$ in a randomly-wired
      (unclustered) network with the same degree distribution (although we also give
       some results for the degree correlations, see section \ref{sect:SFN} and Appendix B).
        Our results
      indicate that for a given level of clustering within this class of structured random networks,
       $p_c$ may be greater than, or less than, $p_c^\text{rand}$, depending of the degree distribution of the network. This contrasts with the results of~\cite{Serrano06b}, where weakly clustered networks (with $c_k<1/(k-1)$) have $p_c>p_c^\text{rand}$, while in the strongly clustered case with $c_k>1/(k-1)$, the clustering decreases the threshold, so $p_c<p_c^\text{rand}$. Indeed, we show in section~\ref{sect:constantF} that the Trapman model with $f_k=F$, a constant for all $k$, leads to clustering increasing the percolation threshold: $p_c>p_c^\text{rand}$, whereas the classification of this case as strongly clustered according to~\cite{Serrano06b} (since $c_k=F(1-2/k)$ here) would predict the opposite conclusion.

Similarly, Figure~\ref{f:3} gives clear examples of cases (e.g. $\beta=2$) where $c_k<1/(k-1)$, but the result of $p_c<p_c^\text{rand}$ is the opposite to that predicted by~\cite{Serrano06b} for the weakly clustered case. These contradictions to the results of~\cite{Serrano06b} are not surprising when we consider that the approach of~\cite{Serrano06b} is focussed on clustering due to loops of length three (i.e. triangles) in the graph. Indeed, the authors of~\cite{Serrano06b} carefully point out that they do not consider effects of longer loops. By contrast, the clustering within the Trapman model is more heavily localized, since a node of degree $k$ which is a member of a triangle must also be part of a loop of length $n$ for all $n$ from $3$ to $k$. Therefore we should not expect the theory of~\cite{Serrano06b} to apply to the Trapman model; nevertheless it is instructive to find that model networks with the same degree distributions $P_k$ and clustering coefficients $c_k$ can give opposite results for this important question. Higher order information, e.g. some measure of the density of loops of length greater than three~\cite{Kim05}, is required to distinguish the two types of networks from each other.

The model of clustering described here has the important advantage
of analytical tractability, permitting us to calculate the bond
percolation threshold and sizes of $k$-cores and giant connected
components. However, the model is limited in its applicability to
real-world networks by the rather artificial structure of
clustering using $k$-cliques, which is not expected to be the
dominant form of triangle-formulation within most real-world
networks. Bearing in mind this caveat, we use the $P_k$ and $c_k$
parameters of some real-world networks (see Table~\ref{table1}) to
find the values of $p_c$ predicted by equation~(\ref{GCCcond}). In
some cases (power grid, PGP) we find $p_c>p_c^\text{rand}$, while
in others (e.g. Internet, WWW) the opposite conclusion is reached.
The applicability of this and related models to real-world
networks will be the topic of further study.

\begin{acknowledgments}
This work was funded by Science Foundation Ireland under programmes 06/IN.1/I366, MACSI, and 05/RFP/MAT0016.
\end{acknowledgments}

\appendix
\section*{Appendix A: Clique calculations}
Newman~\cite{Newman03b} gives results relevant to the bond percolation problem on a $k$-clique, i.e. a complete graph of $k$ nodes. Here we briefly review these results and show they can be applied to calculate connectivity properties of the individuals graph.

For bond occupation probability $p$, the damaged $k$-clique may consist of a number of disconnected clusters of nodes. Letting $P(m|k)$ be the probability that a randomly chosen node in the damaged $k$-clique belongs to a connected cluster of $m$ nodes (including itself), equation~(7) in~\cite{Newman03b} gives
\begin{equation}
P(m|k) = \lb(\!\!\begin{array}{c}
                k-1 \\
                m-1
                \end{array}\!\!\rb) (1-p)^{m(k-m)}P(m|m).
\end{equation}
The probabilities $P(m|m)$ may be determined iteratively from the relation
\begin{equation}
P(k|k) = 1-\sum_{m=1}^{k-1} P(m|k),
\end{equation}
with $P(1|1)=1$. Consider an individual $A$ in a damaged household of $k$ individuals. We seek the number of external super-individuals which are connected to $A$ via undamaged paths through his household---note we do not count $A$'s own direct external link. The individual $A$ is connected to $m-1$ other individuals in the household with probability $P(m|k)$, and each of these other individuals has a single link external to the household, which is undamaged with probability $p$. Thus the average number of undamaged external links from the connected cluster (and hence from $A$) to other super-individuals is
\begin{equation}
D_k(p)=p\sum_{m=1}^k (m-1) P(m|k).
\end{equation}
The polynomials $D_k(p)$ for some low values of $k$ are given
below:
\begin{align}
D_3(p) = 2 p^2 (&1+p-p^2) \nn\\
D_4(p) = 3 p^2 (&1+2p-7p^3+7p^4-2 p^5) \nn\\
D_5(p) = 4 p^2 (&1+3 p+3 p^2-15 p^3-27 p^4+127 p^5 \nn\\
 &-175 p^6+120 p^7 -42 p^8 + 6 p^9).
\end{align}

\section*{Appendix B: Degree-degree correlations}
We consider the calculation of $P(k,j)$, the joint pdf of degrees
of vertices at either end of a randomly chosen edge in the
individuals graph, for the special case of $f_k=F=1$ for all $k$,
and with $P_k=0$ for $k<3$. We begin by noting that the number of
edges in the super-graph is $\widetilde N \widetilde{z}/2$, and
each of these also exists in the individuals graph as an
\emph{external edge} joining two individuals in different
households. Since $F=1$, every super-individual of degree $k$ is a
household, and so is expanded in the individuals graph to a
$k$-clique---this adds a total of $\widetilde N \sum_k \wP_k
k(k-1)/2$ further edges to the individuals graph. Therefore, a
randomly chosen edge in the individuals graph is an external edge
with probability
\begin{equation}
\frac{\wz}{\wz+\sum_k \wP_k k (k-1)} = \frac{\wz}{\sum_k \wP_k
k^2},
\end{equation}
and using equation (\ref{pp}) with $f_k\equiv 1$ (and $P_k=0$ for
$k<3$) reduces this to $1/z$.

An external edge has end-vertex degrees $k$ and $j$ with
probability
\begin{equation}
\frac{k \wP_k}{\wz}\frac{j \wP_j}{\wz}= P_k P_j,
\end{equation}
since the super-graph is an uncorrelated random graph. An internal
edge is in a $j$-clique with relative probability
\begin{equation}
\frac{\wP_j j (j-1)/2}{\sum_{\kp}\wP_\kp \kp(\kp-1)/2} =
\frac{(j-1)P_j}{z-1}
\end{equation}
and its end-vertex degrees are both equal to $j$. Combining all
the possibilities, we obtain equation~(\ref{Pkj}):
\begin{equation}
P(k,j) = \frac{1}{z} P_k
P_j+\lb(1-\frac{1}{z}\rb)\frac{(j-1)P_j}{z-1} \delta_{k j}.
\end{equation}
The average degree of neighbors of nodes with degree $k$ is then
\begin{eqnarray}
\lb<k\rb>_{\text{nn}} &=& \frac{\sum_j P(k,j)j }{\sum_{j} P(k,j)}
\nonumber
\\& = & k-1+\frac{z}{k}.
\end{eqnarray}

\section*{Appendix C: Relation between order parameter and edge statistics }
Following \cite{Goltsev06}, we derive here equation (\ref{eqn35}) for the fraction of edges in the super-graph which link two unpruned super-individuals, i.e. super-individuals belonging to the $K$-core. Note from the discussion preceding equation \eqref{Fh} that all individuals of a household are in the same state and so we may speak of super-individuals as pruned or unpruned.

Consider the super-graph where the cascade has ended and all the nodes in the graph have been updated. Let us first calculate $L_K$, the number of edges in the super-graph which connect unpruned super-individuals. Taking all super-individuals one by one and counting links to any of their unpruned neighbors (if the chosen super-individuals is itself unpruned) will give $2 L_K$.

In order to calculate the expected value of this quantity we consider a randomly-chosen super-individual of the super-graph. Taking this as the root of the tree approximation of the super-graph, we suppose it has degree $k$ and $m\le k$ pruned children. The probability that $m$ of its $k$ children are pruned (meaning that $k-m$ children are unpruned) is $\binom{k}{m} q_\infty^m(1-q_\infty)^{k-m}$, where $q_\infty$ is the order parameter given by the solution of the self-consistent equation (\ref{SCK}).

The state of the root depends on the state of its children as follows. The root can be either a bachelor (which happens with probability $1-g_k$) or a household (which happens with probability $g_k$). In each of these cases it is respectively pruned with probability $F_b(m,k)$ and $F_h(m,k)$, which are given by equations~\eqref{Fb} and~\eqref{Fh}. Therefore, the probability that the root chosen at random is pruned when it has $m$ pruned children is given by a weighted sum of probabilities
\begin{align}
 \widetilde F(m,k) &= (1-g_k) F_b(m,k) + g_k F_h(m,k)\\
\nn &= W_k F_b(m,k),
\end{align}
where $W_k$ is defined by equation \eqref{Wk}.


Combining the probabilities together, the expected number of edges linking an unpruned root of degree $k$ to its unpruned children is
\begin{equation}
\sum_{m=0}^k (k-m) \binom{k}{m} q_\infty^m(1-q_\infty)^{k-m} \lb(
1 - \widetilde{F}(m,k) \rb), \label{C1a}
\end{equation}
where the $(k-m)$ factor counts the unpruned children, given that
$m$ of the $k$ children are pruned, while the
$(1-\widetilde{F}(m,k))$ term accounts for the root node being
unpruned. Averaging this over the degree distribution of the
super-graph and multiplying by the number of nodes  gives
\begin{align}
\nonumber &2L_K = \\
&\widetilde{N} \sum_{k=0}^\infty \wP_k \sum_{m=0}^k (k-m) \binom{k}{m} q_\infty^m(1-q_\infty)^{k-m} \lb( 1 - \widetilde{F}(m,k) \rb).
\label{C1}
\end{align}
The fraction of edges in the super-graph linking unpruned
super-individuals is found by dividing the right hand side of
(\ref{C1}) by the total number of edges in the super-graph $L =
\widetilde{N} \wz/2$, to obtain
\begin{align}
\nn &\frac{L_K}{L} = \\
& \sum_{k=0}^\infty \frac{\wP_k}{\wz} \sum_{m=0}^k (k-m)
\binom{k}{m} q_\infty^m(1-q_\infty)^{k-m} \lb( 1 - \widetilde{F}(m,k) \rb).
\end{align}
Using the identity $(k-m)\binom{k}{m} = k \binom{k-1}{m}$ and
factoring out $(1-q_\infty)$, this can be written as
\begin{eqnarray}
&&\hspace{-0.2cm}(1-q_\infty)\times \nonumber\\
&&\hspace{-0.5cm} \sum_{k=0}^\infty \frac{k \wP_k}{\wz}
\sum_{m=0}^{k-1}  \binom{k-1}{m} q_\infty^m(1-q_\infty)^{k-1-m}
\left( 1 - \widetilde{F}(m,k) \right). \nonumber\\
\hspace{-0.2cm}
\end{eqnarray}
Finally, rewriting the last expression as
\begin{eqnarray}
&&\hspace{-0.2cm}(1-q_\infty)\times \nonumber\\
&&\hspace{-0.5cm}\left( 1 - \sum_{k=0}^\infty \frac{k \wP_k}{\wz}
\sum_{m=0}^{k-1}  \binom{k-1}{m} q_\infty^m(1-q_\infty)^{k-1-m}
\widetilde{F}(m,k)\right),
\nonumber\\
\hspace{-0.2cm}
\end{eqnarray}
and using (\ref{SCK}) gives  $(1-q_\infty)^2$. Equation
(\ref{eqn35}) of the main text follows immediately.



\section*{Appendix D: Zero-clustering limit of K-core size}
In the unclustered case the self-consistent equation (\ref{SCK})
reduces to $q_\infty = H(q_\infty)$, with $H(q_\infty)$ given by
\begin{equation}
\sum_{k=1}^\infty \frac{k}{z}P_k \sum_{m=0}^{k-1} \binom{k-1}{m}
q_\infty^m (1-q_\infty)^{k-1-m} F_b(m,k),
\end{equation}
where $F_b(m,k)=1$ if $m>k-K$ and zero otherwise. We show that the
right-hand side of this equation is the same as in equation (2) of
\cite{Dorogovtsev06} in the undamaged  networks case.

The sum over $k$ is first expressed as a sum over $i$, with
$i=k-1$:
\begin{equation}
\sum_{i=0}^\infty \frac{(i+1)}{z}P_{i+1} \sum_{m=0}^{i}
\binom{i}{m} q_\infty^m (1-q_\infty)^{i-m} F_b(m,i+1).
\end{equation}
Next, the sum over $m$ is re-ordered to a sum over $n$, with
$n=i-m$, and using the fact that $\binom{i}{m} = \binom{i}{n}$:
\begin{equation}
\sum_{i=0}^\infty  \sum_{n=0}^{i} \frac{(i+1)}{z}P_{i+1}
\binom{i}{n} q_\infty^{i-n} (1-q_\infty)^{n} F_b(i-n,i+1).
\end{equation}
The double sum $\sum_{i=0}^\infty \sum_{n=0}^i $ can be rewritten
as $\sum_{n=0}^\infty\sum_{i=n}^\infty$, and using the fact that
$F_b(i-n,i+1)$ is 1 only for $n<K-1$ we obtain
\begin{equation}
\sum_{n=0}^{K-2}  \sum_{i=n}^{\infty} \frac{(i+1)}{z}P_{i+1}
\binom{i}{n} q_\infty^{i-n} (1-q_\infty)^{n}.
\end{equation}
This, with the notation mapping $q_\infty \mapsto R$, gives
equation (2) of \cite{Dorogovtsev06} (with $p=1$). Similar
manipulations reduce the zero-clustering version of equation
(\ref{rhoeqn}) to equation (1) of \cite{Dorogovtsev06}, with the
notation mapping $\rho \mapsto 1-M$.
\bibliography{networks}
\end{document}